\documentclass[preprint,5p,times,twocolumn,authoryear]{elsarticle}

\usepackage{lineno,hyperref}

\usepackage[cmex10]{amsmath}
\usepackage{amssymb}
\usepackage{mathrsfs}
\usepackage{amsthm}
\usepackage{lineno}
\usepackage{graphicx}
\usepackage{color}
\usepackage{algorithm}
\usepackage{algpseudocode}

\usepackage[algo2e,ruled,linesnumbered]{algorithm2e}
\usepackage{lipsum}

\usepackage{bm}
\usepackage{setspace}
\usepackage{hyperref}
\DeclareGraphicsExtensions{.pdf,.png,.jpg,.eps,.ps}

\def\ba{\begin{array}}
	\def\ea{\end{array}}
\def\baa{\begin{align}}
	\def\eaa{\end{align}}

\newcommand{\bsq}{\begin{subequations}}
	\newcommand{\esq}{\end{subequations}}

\newcommand{\beq}{\begin{equation}}
\newcommand{\eeq}{\end{equation}}
\newcommand{\bq}{\begin{eqnarray}}
\newcommand{\eq}{\end{eqnarray}}
\newcommand{\bqn}{\begin{eqnarray*}}
	\newcommand{\eqn}{\end{eqnarray*}}
\newcommand{\bee}{\begin{enumerate}}
	\newcommand{\eee}{\end{enumerate}}
\newcommand{\bi}{\begin{itemize}}
	\newcommand{\ei}{\end{itemize}}

\newcommand{\diag}{\mathrm{diag}}

\usepackage{comment}
\usepackage{ifthen}
\newboolean{showcomments}
\setboolean{showcomments}{true}
\newcommand{\wang}[1]{\ifthenelse{\boolean{showcomments}}
	{ \textcolor[rgb]{1,0,1}{(ZW:  #1)}}{}}
\newcommand{\fliu}[1]{\ifthenelse{\boolean{showcomments}}
	{ \textcolor{red}{(FL:  #1)}}{}}
\newcommand{\slow}[1]{\ifthenelse{\boolean{showcomments}}
	{ \textcolor{blue}{(SL:  #1)}}{}}

\newtheorem{theorem}{Theorem}
\newtheorem{lemma}[theorem]{Lemma}

\newtheorem{proposition}[theorem]{Proposition}

\newtheorem{definition}{Definition}
\newtheorem{remark}{Remark}

\newtheorem{example}{Example}
{}

\allowdisplaybreaks

\journal{xxx}

\begin{document}
	
\graphicspath{{Paper_Fig/}}
%\setstretch{0.96}

\begin{frontmatter}
	\title{Structure and Control of Biology-inspired Networks }

	\author[us]{Zexin~Sun\corref{mycorrespondingauthor}}\ead{zxsun@bu.edu}
	\cortext[mycorrespondingauthor]{Corresponding author}
	\author[usa]{John~Baillieul}\ead{johnb@bu.edu}
              % e-mail address   % (ead) as shown
\address[us]{Division of Systems Engineering at Boston University, MA 02215, USA}  % Please supply                                              
\address[usa]{Departments of Mechanical Engineering, Electrical and Computer Engineering, and the Division of Systems Engineering at Boston University, MA 02215, USA}

\begin{abstract}
%Utilizing contemporary progress in the fields of neuroscience and control theory, the present paper proposes a biologically inspired network model featuring dynamic connections regulated by (anti-)Hebbian learning principles. Formal examination rooted in graph theory and classical control proves that this biologically plausible model displays boundedness, stability, and structural controllability (observability) contingent on a generalized cactus configuration with multiple control nodes. In addition, the proposed model demonstrates resilience to changes in network connections and cascade connections without loss properties when the generalized cactus structure is preserved. Simulations employing both a toy example and the visual system of the macaque monkey with various input types corroborate the model's efficacy in reproducing fundamental neural characteristics.
There is increasing interest in developing the theoretical foundations of networked control systems that illuminate how brain networks function so as to enable sensory perception, control of movement, memory and all the operations that are needed for animals to survive.  The present paper proposes a biologically inspired network model featuring dynamic connections regulated by Hebbian learning.  Drawing on the machinery of graph theory and classical control we show that our novel nonlinear model exhibits such biologically plausible features as bounded evolution, stability, resilience, and a kind of structural stability --- meaning that perturbations of the model parameters leave the essential properties of the model in tact.  The proposed network model involves generalized cactus graphs with multiple control input nodes, and it is shown that the properties of the network are resilient to various changes in network topology provided these changes preserve the generalized cactus structure.  A particular example described in what follows is an idealized network model of the visual system of a macaque monkey.  The model displays resilience to network disruptions such as might occur in a living organism due to disease or injury.  A different model of the same type provides an example of a system that can perform data classification.

\end{abstract}

\begin{keyword}
	Neuromimetic Networks; 
	Hybrid system;
	Structural controllability;
	Cactus structure.
\end{keyword}
\end{frontmatter}

\section{Introduction}
Animals demonstrate complex behaviors, spanning basic motor functions (e.g., limb movements) to cognitively demanding predictive tasks, emergent from the bursty interaction dynamics within distributed neuronal ensembles. An interdisciplinary corpus of research, as surveyed in \cite{gauthier2010perceptual}, has begun elucidating the links between underlying neural architectures and the perceptual and behavioral correlates of measured brain activity across phylogenetically diverse species. The neurophysiology of the animal kingdom exhibits substantial complexity, with systems comprising as few as 302 neurons, interconnected by 460 synapses in rudimentary nematodes \cite{cook2019whole}, scaling up to an estimated 86 billion neurons and $3 \times 10^{14}$ synaptic connections in an adult human brain \cite{herculano2009human}. Nevertheless, even simple organisms like {\em C. elegans} manifest varied life processes orchestrated by neurophysiological dynamics that remain incompletely understood.  Decades of study of biological neural systems has inspired the development of artificial neural networks and techniques that allow them to learn.  Among the most celebrated of these, backpropagation has had notable success enabling ANN's in applications to speech recognition, machine vision, and natural language processing. Artifical neural networks are fundamentally different from biological neural networks, however.  While cortical feedback connections are demonstrably pervasive anatomically, prevailing conceptualizations of the backpropagation algorithm do not exhibit a biologically plausible mechanism by which the connections in the cortex could convey error signals required for synaptic modification under a strict supervision paradigm \cite{lillicrap2020backpropagation}. Indeed, \cite{michel2022survey} have studied backpropagation-trained models, which display acute vulnerability, with even minor perturbations of inputs invoking significant fluctuations in output. Relative to the resilience of human cognitive capacities, this divergence underscores pronounced constraints within deep network architectures regarding the capacity for generalization and conceptual abstraction. %\cite{rumelhart1985learning} proposed feedforward architectures with feedback loops yet achieve comparable performance on pattern recognition benchmarks.
At the same time, \cite{krotov2019unsupervised} have underscored the differences between supervised learning with backpropagation and the biologically plausible Hebbian type of network training in which synapse strengths are adjusted locally and depend only on the activities of the pre- and postsynaptic neurons.  Interestingly, the Hebbian model of \cite{krotov2019unsupervised} has performed comparably to standard models trained by backpropagation on the benchmark MNIST and CIFAR-10 datasets. Nevertheless, significant open questions persist regarding model generalizability, scaling, and applicability to real-world domains. Formulating neurobiologically inspired frameworks remains an important research direction aimed at assimilating core principles from biological neural systems into enhanced machine learning methodologies. With this as motivation, the present paper considers a biologically plausible dynamic system  with recurrent connections governed by (anti-)Hebbian learning rules that learn weights locally and accept the inputs sequentially. We leverage tools from control and graph theory to find conditions to guarantee its stability and robustness properties.\\
\\
\textit{Literature Review}
A brief background survey of bio-inspired and neuro-inspired control theory could divide current research into two broad areas: the micro case and the macro case. The micro case considers systems with a large number of simple control inputs that are coordinated to achieve prescribed actions. In this context, \cite{beyeler2019neural} viewed such simple inputs as neuronal responses, which can be understood as an emergent property of nonnegative sparse coding (NSC). %NSC might be utilized by sensory cortical regions to efficiently encode representations of external stimuli, while basal ganglia may employ NSC for the coordination of motor behaviors. 
Building upon this notion of simple neuronal inputs, \cite{baillieul2019perceptual} and \cite{sun2022neuromimetic} proposed {\em neuromimetic} control models that utilize such inputs collectively, associating them with system-wide objectives.  %\cite{sun2022neuromimetic} studied the dynamic characteristics of control systems actuated by the aggregate effects of inputs — simulating neural states of spiking or resting. 
Following this investigation, \cite{sun2024neuromimetic} explored neural heuristic quantization and showed that quantized control input sequences acting in parallel through large numbers of input channels could achieve control objective while being resilient to channel dropouts. %It was demonstrated that by employing algorithms akin to Oja's rule by \cite{oja1982simplified}, these neuromimetic models could emulate classical control designs produced by  analog inputs. 
This line of research primarily focuses on how multiple neurons collaborate to accomplish tasks, closely resembling real microscopic biological models. However, it has certain limitations. For instance, when constructing control systems, directly integrating biologically plausible principles into system parameters proves challenging, often requiring the reconstruction of algorithms as a supplement for simultaneous application. Furthermore, while this approach is effective for tasks such as emulation, it faces difficulties in addressing a broader range of real-world challenges, including complex problems like data classification, path planning, and others.

The macro case, as treated in this paper, investigates biologically plausible implementations coupled with different learning rules, such as the Hebbian rules introduced by \cite{galtier2012hebbian}, which were shown to be capable of learning geometric features of cortical organization by extracting the geometry of input spaces using slow Hebbian learning in Hopfield networks, and Oja's rules explored by \cite{pehlevan2015hebbian}, which proposes a single-layer neuronal network that performs online symmetric non-negative matrix factorization (SNMF) of the similarity matrix of streamed data.
The algorithm is derived from a principled cost function and can be implemented by a biologically plausible network with local learning rules. %An extension work by \cite{pehlevan2019neuroscience} demonstrated that neural networks equipped with both Hebbian and anti-Hebbian learning rules can perform a broad range of unsupervised learning tasks. 
Meanwhile, \cite{krotov2019unsupervised} have proposed a novel biologically plausible learning rule that utilizes global inhibition in the hidden layer to extract key features, demonstrating its efficacy on the MNIST dataset. While all of the aforementioned papers are grounded in Hebbian learning and recurrent dynamics, each adopts a specific mathematical framework, such as SNMF, rather than providing a more comprehensive biological description. This methodological focus introduces certain limitations, as backpropagation algorithms may be replaced in initial stages, yet the scope of tasks that can be effectively addressed remains restricted. In some cases, this approach necessitates the addition of another layer trained with labeled data using conventional methods, such as stochastic gradient descent, as discussed by \cite{krotov2019unsupervised}.Furthermore, these studies generally lack an investigation into critical properties such as the controllability and stability of the system. Recently, a similar work with this paper by \cite{centorrino2024modeling} provided an overview of the biologically inspired models and combined the Hopfield Neural Network and Firing Rate Neural Network with two different Hebbian learning rules, analyzing the contractivity of each coupled model by leveraging non-Euclidean contraction arguments. While it primarily focuses on theoretical analysis without presenting practical application scenarios, our work extends beyond theoretical exploration. By leveraging graph theory and control theory, we analyze the system's various biologically plausible properties. Additionally, we demonstrate the applicability of our approach through its implementation on a macaque visual neural network from \cite{kandel2000principles} and its use in data classification tasks.\\%Indeed, when analyzing dynamic models, a crucial problem that arises is ensuring the stability and controllability of the system
\\
\textit{Contributions}

In the context of the above literature, %the following sections propose network-based models aimed at reproducing biological capabilities such as learning and predictive cognition. The frameworks focus both on individual neurons and on interconnected brain regions.
this paper offers four key contributions: (i) we propose a coupled pre- and post-synaptic model undergoing Hebbian learning and anti-Hebbian rules, capturing state-dependent synaptic plasticity in both excitatory and inhibitory connections. The updating rules for the weights of synapses use only local information based on the neuron connections;  (ii) we leverage graph theory and control theory to analyze this model and derive conditions ensuring reflection of neuronal characteristics, such as stability, structural controllability and resilience; (iii) we give sufficient conditions based on biologically meaningful quantities, such as synaptic decay rate and topological connectivity that preserve neural properties;
(iv) finally, we validate theoretical results via simulations that include the dynamics of a model monkey brain parcellation.

The remainder of this paper is organized as follows. Section 2 provides preliminary notations and motivations for the proposed model. Section 3 describes the model on which the research has been based. Section 4 presents formal analyses of the model features: stability, structural controllability(observability), nonlinearity, and resilience of neural connections. Section 5 demonstrates via simulations that the model exhibits biologically plausible neural dynamics. Finally, Section 6 summarizes the key contributions and results.

\section{Preliminaries}
We denote by $\mathbb{R}$ the set of real numbers and $|| \cdot||_\infty$ the infinity norm. Define the clipping function $[y]_{\underline{y}}^{\overline{y}}=\begin{cases}
	\overline y,& \text{ if } \overline y<y \\
	y,& \text{ if } \underline y\le y\le \overline y\\
	\underline y,& \text{ if } y<\underline y
\end{cases}$.
The symbol $``\circ"$ denotes Hadamard product and $``\preceq"$ denotes element-wise comparisons of the matrix entries. $sign(\cdot)$ is the sign function that has the value +1, -1 or 0 according to whether the sign of a given real number is positive or negative, or zero. The difference of the set $\mathcal{P}$ and $\mathcal{U}$ is defined as the set comprising elements belonging to $\mathcal{P}$ excluding any elements contained in $\mathcal{U}$. Symbolically, this is denoted $\mathcal{P}-\mathcal{U}$.

\subsection{Graph theory and Structure Patterns of Linear Systems}
Let $\mathcal{G}=(\mathcal{V},\mathcal{E})$ denote a directed graph, where $\mathcal{V}$ is the node set and $\mathcal{E}\subseteq \mathcal{V}\times\mathcal{V}$ is the edge set. $N=|\mathcal{V}|$ denotes the total number of nodes. In addition, the edges with positive and negative weights are separated into $\mathcal{E}^+$ and $\mathcal{E}^-$ with $\mathcal{E}^+\cup\mathcal{E}^-=\mathcal{E}$. We recall that the weight matrix of a labeled graph is denoted by $A=[a_{ij}]$, where $a_{ij}$ represents the edge weight from node $j$ to $i$. Define $d^{in}$ as the maximum in-degree of the digraph $\mathcal{G}$. The union and intersection of two graphs, $\mathcal{G}_a$ and $\mathcal{G}_b$, are represented as $\mathcal{G}_a \cup \mathcal{G}_b$ and $\mathcal{G}_a \cap \mathcal{G}_b$, respectively. We denote $\mathcal{G}_a$ as a subgraph of $\mathcal{G}_b$, written as $\mathcal{G}_a \subset \mathcal{G}_b$, if all nodes and edges of $\mathcal{G}_a$ are subsets of the node set and edge set of $\mathcal{G}_b$.

A directed {\em path} in a graph is a sequence of edges such that the terminal node of each edge coincides with the initial node of the subsequent edge, and no node is repeated as both an initial and terminal node within the sequence, excepting the first and last nodes. If the initial node and the terminal node of the path are allowed to coincide, the sequence of edges is called a \emph{cycle}. %The path together with its traversing nodes is referred to as a stem. %Analogously, the cycle together with its traversing nodes is referred to as a bud.
\begin{comment}
   \begin{definition}
	\label{definition_stem}
	(Stem) For a set of nodes $\{n_1,n_2,\dots,n_T\}$, if there are only oriented edges for every $(n_i,n_{i+1})$ pair for all $i=1,2,\dots T-1$, such a sequence of edges and nodes is called a \emph{stem}.
\end{definition} 
Fig. \ref{fig:stem} illustrates a diagram of a stem (in grey) with two bidirectional buds attached to it.
\begin{figure}[h]
	\begin{center}
		\includegraphics[scale=0.35]{stem.png}
	\end{center}
	\caption{ The diagram shows a gray stem $S$ with two orange buds $B_1,B_2$ detached from it. The joints both on the distinguished edges and the stem are denoted by $n_{b_1}$ and $n_{b_2}$.}
	\label{fig:stem}
\end{figure}
\end{comment}

\begin{figure}[h]
	\begin{center}
		\includegraphics[scale=0.65]{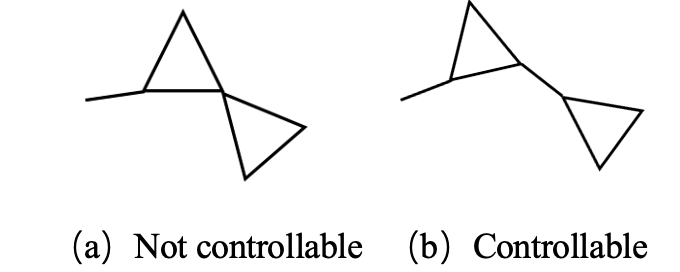}
	\end{center}
	\caption{ The figure illustrates two general cactus structures as described by \cite{arcak2011diagonal}, with the structure on the right also exemplifying Lin's concept of a (spanned) cactus graph.}
	\label{fig:cactus}
\end{figure}

In the present paper, we shall be interested in a special class of graphs known as \emph{cactus graphs}.  %These have proven to be useful in control and network theory in a number of ways---particularly as related to dissipativity and stability.  
We refer to \cite{arcak2011diagonal} for a review of recent literature, which defines cactus graphs as structures where any two simple cycles share at most one common vertex, as illustrated in Fig. \ref{fig:cactus}. However, it is important to note that such a structure alone does not constitute a sufficient condition for graph controllability. For instance, Fig. \ref{fig:cactus}(a) depicts a graph that is not controllable, whereas Fig. \ref{fig:cactus}(b) represents a controllable graph. Consequently, our subsequent work builds upon and significantly expands the research initiated by \cite{lin1974structural}. To further develop these concepts, we adopt Lin's definition of cactus graphs as a foundational framework for our analysis.

\begin{definition}
	\label{definition_bud}
	(Bud) A \emph{bud} is a subgraph containing a cycle and an additional node that is connected by a single edge to a node in the cycle. We call this connecting edge the \emph{distinguished edge} and the additional node \emph{joint node}.
 %A \emph{bud} is a subgraph consisting of a cycle and a node outside the cycle with a single edge connected to the cycle, where the edge is called a distinguished edge.
	%A input(output)-perspective \emph{bud} is formed by a circle $R$ and a distinguished edge $e$, where the terminal(initial) node of $e$ is contained in $R$, but the initial(terminal) node is not. We call such a node to be a joint.
\end{definition}

A number of authors, including \cite{lin1974structural,rech1990structural,mayeda1981structural} have studied buds with directed distinguished edge connections. In these papers, a distinction is made between \emph{input bud} and \emph{output bud}, where the edge is directed from the cycle or the edge is directed toward the cycle. In this paper, we focus on the \emph{bidirectional bud}, which refers to both input and output perspectives, with a bidirectional distinguished edge between the cycle and joint node. Keeping with the botanical terminology, Lin referred to a simple path to which buds are attached as a {\emph stem}.  %For each bud, we refer to the endpoint of the distinguished edge that is not on the cycle as a \emph{joint}, and in Lin's definition of a \emph{cactus graph}, we refer to the nodes located on the stem that connect the bud to it as the \emph{joints}. 

\begin{definition}%\cite{mayeda1981structural}
	\label{definition_cactus}
	(Cactus) Given a stem $S$ and buds $B_1,B_2,\dots,B_n$, the structure $S\cup B_1\cup\dots\cup B_n$ is called a cactus if for all $i\in\{1,2,\dots n\}$, 
 $\{S\cup B_1\cup\dots\cup B_{i-1}\}\cap B_i=v_{b_i}$, where $v_{b_i}$ is the joint node of bud $B_i$.  %(The dangling node being the node on the distinguish edge opposite to the joint.)
\end{definition}
The cactus structure described herein aligns with the definition provided in \cite{lin1974structural} and \cite{menara2018structural}. It is important to note that this conceptualization differs slightly from the definition of a cactus tree in graph theory, where any two simple cycles share at most one common vertex. We say that a graph $\mathcal{G}_a=(\mathcal{V}_a,\mathcal{E}_a)$ is \emph{spanned} by a graph $\mathcal{G}_b=(\mathcal{V}_b,\mathcal{E}_b)$ when $\mathcal{V}_a=\mathcal{V}_b$ and $\mathcal{E}_b\subseteq \mathcal{E}_a$. $\mathcal{G}_a$ and $\mathcal{G}_b$ can be connected or disconnected. 

\begin{definition}
	\label{definition_g-cactus}
	(Generalized cactus) A digraph possesses a generalized cactus structure if it is spanned by a disjoint union of cacti.
\end{definition}

\begin{figure}[h]
	\begin{center}
		\includegraphics[scale=0.67]{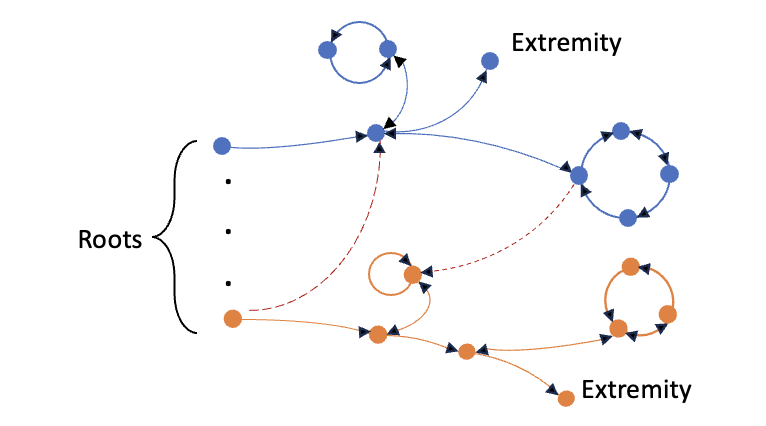}
	\end{center}
	\caption{ A diagram of generalized cactus structure with buds having both directed and bidirectional edges connecting to stems. Red dotted lines denote connections between nodes from two different cacti.}
	\label{fig:cacti}
\end{figure}

In a generalized cactus structure, the root nodes of cacti are the initial nodes of their stems, while the terminal endpoints of stems are called extremities of the cacti. The component cacti may or may not be connected. Fig. \ref{fig:cacti} is an example of a generalized cactus.

\begin{comment}
	
	A digraph is said to be symmetric if it has the property that for every pair of nodes $i,j$, if $(i,j)\in\mathcal{E}$, then $(j,i)\in\mathcal{E}$ and $a_{ij}=a{ji}$. In addition, symmetric digraphs also have similar definitions of cactus and cacti.
	\begin{definition}\label{definiton4}
		(Sym-cactus and Sym-cacti) \cite{menara2019} A sym-cactus is a connected digraph $\mathcal{G}$ with symmetric weights and defined as $\mathcal{G}={\textstyle \overline{\bigcup}_{i=1}^{m}} \mathcal{G}_i$, where $\mathcal{G}_i$ is symmetric cycle or a single node and $\mathcal{V}_i\cap\mathcal{V}_j=\emptyset, \forall i\neq j$. A symmetric digraph $\mathcal{G}$  is called sym-cacti if it is spanned by a disjoint union of sym-cactus.
	\end{definition}
\end{comment}

In classical control theory, the controllability and observability of a linear time-invariant dynamical system $\left\{\begin{matrix}
	\dot x=&\hat Ax+\hat Bu \\
	y=&\hat Cx
\end{matrix}\right.$ can be determined by examining the rank of the {\em controllability matrix} $[\hat{B},\hat{A}\hat{B},\dots, \hat{A}^{N-1}\hat{B}]$ and the {\em observability matrix} $ [\hat{C};\hat{C}\hat{A};\dots; \hat{C}\hat{A}^{N-1}]$. However, this rank criterion is often too strict, as the coefficient matrices $\hat{A}$, $\hat{B}$ and $\hat{C}$ are frequently unknown, and measurements may be imprecise. Thus, \cite{lin1974structural} proposed that a linear time-invariant system is \emph{structurally controllable} if the controllability matrix is full rank for almost all choices of non-zero entries of $\hat{A}$ and $\hat{B}$. To pursue the concept of structural controllability, we introduce the notion a {\em matrix structure}.  The idea going back to \cite{lin1974structural} is that zero entries in matrices play a distinguished role.  We say that two matrices have the same structure if they have the same dimensions and zero entries in the same locations---with arbitrary non-zero entries elsewhere.
\begin{example}
	Consider example strucrture patterns of $\mathcal{A}$ and $\mathcal{B}$.
	$\mathcal{A}=\begin{bmatrix}
		*& * &0 \\
		0& 0 &* \\
		*&0  &0
	\end{bmatrix}$,  $\mathcal{B}=\begin{bmatrix}
		0& 0 &0 \\
		*& 0 &0 \\
		0&0  &0
	\end{bmatrix},$
	where $*$ denote the nonzero entries. If $\mathcal{A}_1=\begin{bmatrix}
		1& 2 &0 \\
		0& 0 &2 \\
		1&0  &0
	\end{bmatrix}$, $\mathcal{B}_1=\begin{bmatrix}
		0& 0 &0 \\
		1& 0 &0 \\
		0&0  &0
	\end{bmatrix}$ and $\mathcal{A}_2=\begin{bmatrix}
		2& 1 &0 \\
		0& 0 &3 \\
		2&0  &0
	\end{bmatrix}$, $\mathcal{B}_2=\begin{bmatrix}
		0& 0 &0 \\
		2& 0 &0 \\
		0&0  &0
	\end{bmatrix}$, the pairs $(\mathcal{A}_1,\mathcal{B}_1)$ and $(\mathcal{A}_2,\mathcal{B}_2)$ have the same structure pattern and we say that both belong to the structure pair $(\mathcal{A},\mathcal{B})$.
\end{example}

\begin{definition}%\cite{lin1974structural}
	\label{definition_sc}
	(Structural Controllability) A linear system $\dot x=Ax+Bu$ is {\em structurally controllable} if there exists a controllable pair $(\hat A,\hat B)$ with the same structure as $(A,B)$.
\end{definition}

\begin{proposition}
    If a system $(A,B)$ is uncontrollable but is structurally
controllable, then almost all perturbations of the matrix entries — $(A+\Delta A,B+\Delta B)$ yield a controllable system.
\end{proposition}

\begin{comment}
{\color{blue}

Perhaps it would be better to formulate the definition of structural controllability in terms of the existence of controllable perturbations arbitrarily close to the system in question.  If this is done, Definition 5 would become a proposition.  The following would then be of interest.

\begin{example}
Consider single input systems of the form $\dot x=Ax+bu$ where $A=\diag{(s_1,\dots,s_n)}$ and $b={\bf 1}$ is the column vector whose entries are all $1$.  It is easy to show that this simple system is controllable if and only if $s_i\ne s_j$ for $i\ne j$.  Hence, if $A=I$, the $n\times n$ identity matrix, the system is not cotrollable.  It is structurally controllable, however.
\end{example}

We should probably demonstrate awareness of "Nodal Dynamics, Not Degree Distributions, Determine the Structural Controllability of Complex Networks" in PLoS ONE.}

\medskip
\end{comment} 

\begin{comment} 
\parbox{2.3in}{\color{blue} Put a Subsection here that formally defines ``system graph'' and how it is related to structure patterns:}
\smallskip

\end{comment}
\subsection{System Graphs and Structure Patterns}
We introduce the concept of a {\em structurally controllable graph} as a topological invariant associated with structurally controllable systems.

\begin{definition} 
Given a linear system $\bm{\dot x}=A\bm{x}+B\bm{u}$, with $\bm{x}\in\mathbb{R}^n,\bm u\in\mathbb{R}^m$, we associate a rooted graph (called a {\em control graph}) with node (vertex) set ${\cal V}=\{v_1,\dots,v_n\}$ and root set ${\cal U}=\{u_1,\dots,u_m\}$.  Ordered pairs of elements in ${\cal V}\times {\cal V}$ correspond to nonzero entries in $A$ with the values $a_{ij}$ serving as edge labels, and ordered pairs in ${\cal U}\times {\cal V}$ correspond to entries in $B$, with values $b_{ij}$ serving as input labels.
\end{definition}
There is a natural connection relating control graphs, structure patterns, and structurally controllable systems.
\begin{definition}
\label{definition_scg}
A control graph is {\em structurally controllable} if any linear system in terms of which it is defined as above is controllable.
\end{definition}
The definition of \emph{structural observability} is given similarly.
\begin{comment}
The concept of a (structurally) controllable graph, as introduced by \cite{lin1974structural}, is analogous to the notion of a (structurally) controllable system. This definition stipulates that the controllability matrix must possess full rank. The controllability matrix is composed of two components: the weight matrix (A) and the input matrix (B). In the context of graph theory, the weight matrix corresponds to the adjacency matrix of the graph, while the input matrix is constructed from the nodes that receive external inputs. Graphs that share a same pattern structure are classified as similar graphs. 
\begin{definition}%\cite{lin1974structural}
	\label{definition_scg}
	(Structurally Controllable of a Graph)  
 A graph is structurally controllable if there exists a similar graph that is controllable.
\end{definition}
\end{comment}
This definition establishes a crucial link between the structural properties of a graph and its relevance for controllability. It suggests that the controllability of a graph is not solely dependent on its specific numerical parameters, but rather on its underlying structure pattern, which is determined by its control graph.  In turn, the structural controllability of a graph can be determined by showing that it has a generalized cactus structure.
In a parallel development, \cite{rech1990structural} established the dual concept of structural observability. For more detail on the state-of-the-art regarding structural properties of linear systems refer to \cite{dion2003generic}.  The concept of graph controllability will be useful in the following sections where we discuss related nonlinear models of Hebbian learning and plasticity in brain networks.

%By analogy with the (structurally) controllable of a system, the definition of the (structurally) controllable graph is proposed by \cite{lin1974structural}, whose controllability matrix formed by the weight matrix and input matrix having full rank. It was proved that the graph is structurally controllable if and only if the graph is spanned by a cactus. Analogously, \cite{rech1990structural} defined the dual concept for \emph{structurally observable}. In this manner, the controllability and observability of a network can be determined solely from the pattern of connections between nodes, without requiring explicit knowledge of the numerical parameter values.  For more detail on the state-of-the-art regarding structural properties of linear systems refer to \cite{dion2003generic}.  

\section{Modeling of Neuromimetic Dynamic Networks}

In this section, we propose a dynamic Hopfield model of biologically-inspired networks that incorporates the dynamics of individual neurons as well as the dynamic changes of neuronal connectomes under Hebbian and anti-Hebbian learning rules. %which only use local information.
The models describe aspects of brain networks that can be described by directed graphs with temporally evolving edge weights. We construct  graphs $\mathcal{G}=(\mathcal{V},\mathcal{E})$, where $\mathcal{V}$ and $\mathcal{E}$ correspond to brain regions and white matter connections in the macro-scale and neurons and synapses in the micro-scale, respectively.   These interoperating micro- and macro-scale models are given by (\ref{eq:model_1}) as follows:

\begin{subequations}
	\label{eq:model_1}
	\begin{align}
		\dot x_{i}(t)&=-c_nx_{i}(t)+\gamma_{\theta}(\sum_{(j,i)\in\mathcal{E}}a_{ij}(t)x_{j}(t))+ b_iu_{i},\label{1a}\\
		a_{ij}(t)&=\left\{\begin{matrix}\label{1b} 
			\left[c_a^-a_{ij}(\tau p)-\phi(x_i(\tau p)x_j(\tau p))\right]_{\underline a^-}^{\overline a^-},(j,i)\in\mathcal{E}^-\\
			\left[c_a^+a_{ij}(\tau p)+\phi(x_i(\tau p)x_j(\tau p))\right]_{\underline a^+}^{\overline a^+},
   (j,i)\in\mathcal{E}^+
		\end{matrix}\right. 
		%\dot a_{ij}&=-c_aa_{ij}+\phi(x_ix_j),i\neq j, \label{1b}
	\end{align}
\end{subequations}
where $t\in(\tau p,\tau (p+1))$, $p=1,2,\cdots$ and $\tau$ is a time constant denoting the time interval of updating the edge weight. $i,j\in \mathcal{V}$. $c_n>0$ represents the decay rate, and $0<c_a^+,c_a^-<1$ exhibits the synaptic depression property. $\phi(\cdot)$ is the activation function. In models of neurodynamics, a commonly used activation function is the sigmoid function.
In (\ref{1a}), $b_i=0$ means there is no external input injected to node $i$. The entry $a_{ij}$ represents the connection weight of the pre-synaptic neuron $j$ to the post-synaptic neuron $i$. 
The function $\gamma_{\theta}$ is designed to emulate neuronal spiking behavior. It is defined as:
\begin{equation}
\nonumber
\gamma_\theta(x)=\left\{\begin{matrix}
 x, & \text{if } |x|>\theta\\
  0,& \text{otherwise}
\end{matrix}\right.,
\end{equation}
where $\theta$ represents the threshold for neuronal activation. It can also be expressed using the standard $ReLU$ function, $\sigma(x)=\max(x,0)$ as
$\gamma_{\theta}(x)=-\sigma(-x-\theta)+\sigma(x-\theta)+sign(\sigma(x-\theta))\theta+sign(-\sigma(-x-\theta))\theta$. %While various activation functions exist, this serves as an illustrative example.
If $x$ represents the membrane potential of the postsynaptic neuron, the function $\gamma_\theta(x)$ models the instantaneous increase in postsynaptic membrane potential caused by the presynaptic release of vesicles, under the condition that the threshold potential $+\theta/-\theta$ is reached. This models the scenario where presynaptic vesicle release results in a discontinuous jump in postsynaptic potential upon crossing the threshold $\theta/-\theta$, which corresponds to the integrate-and-fire model in \cite{burkitt2006review}.
(\ref{1b}) models the evolution of connectomes of neurons or brain regions occurring on slower timescales than neuronal state changes, consistent with synaptic plasticity processes that strengthen or weaken synapses over time as studied by
\cite{citri2008synaptic}. 
Therefore, over a short interval $\tau$, synaptic strengths are fixed and describe the modulation of faster neural state fluctuations described by $x$. Plasticity updates to $A$ occur discretely every $\tau$ units of time. The evolving positive and negative edge weights describe the excitatory and inhibitory plasticity, which correspond to Hebbian and anti-Hebbian rules, respectively.
Meanwhile, it is assumed that the synaptic strength is bounded: $\underline{a}^-<a_{ij}<\overline{a}^-<0$ for all $(j,i)\in\mathcal{E}^-$ and $0<\underline{a}^+<a_{ij}<\overline{a}^+$ for all $(j,i)\in\mathcal{E}^+$. For example, when the dynamic of $a_{ij}$ for $(j,i)\in\mathcal{E}^+$ evolves to exceed the value $\overline a^+$, it takes the value $\overline a^+$; when it is smaller than $\underline a^+$, it becomes $\underline a^+$. 
This assumption is reasonable in neuroscience since synapses are naturally bounded.

In vector form with an output observer, the model (\ref{eq:model_1}) becomes
\begin{subequations}
	\label{eq:model}
	\begin{align}
		 \bm{\dot x}(t)&=-C_n\bm{x}(t)+\Gamma(A(t)\bm{ x}(t))+ B\bm{u},\,\label{2a}\\
		%\dot A&=-C_aA+\phi(xx^T),
		A(t)&=\left[C_a\circ A(\tau p)+sign(A)\circ\phi(\bm{x}(\tau p)\bm{ x}(\tau p)^T)\right]_{\underline A}^{\overline A},\label{2b}\\
		\bm{ y}(t)&=C\bm{ x}(t),\label{2c}
	\end{align}
\end{subequations}
where $ A=[a_{ij}]\in\mathbb{R}^{N\times N}$ is the weight (adjacency) matrix with all its diagonal entries set to zero and, as in Section 2, ``$\circ$'' denotes Hadamard product.  $\bm{u}=[u_i]^T$ is the external stimulus vector.  In this notation, the lower bound on the weight matrix in (\ref{eq:model_1}) is
\[\underline{A}[i,j]=\begin{cases}
	\underline{a}^-,& \text{ if } (j,i)\in\mathcal{E}^- \\
	\underline{a}^+,& \text{ if } (j,i)\in\mathcal{E}^+\\
	0,& \text{ otherwise }
\end{cases}
\]
with the upper bound
$\overline{A}$ having a similar definition. Patterns of the weight matrix can be described by $sign(A)$ which denotes the positive or negative sign of the edge weights.  With this notation, the entries consist of $-1,0,1$, which never change during the evolution of the model.
Therefore, the weight matrix $A$ is confined to a polytope defined by the element-wise maximum $\overline{A}$ and minimum $\underline{A}$ according to the boundedness specified in $(\ref{1b})$, i.e., $\underline A\preceq A\preceq\overline A$. $B=[b_{k_1}\bm{e}_{k_1},\ b_{k_2}\bm{e}_{k_2},\  \dots,\ b_{k_n}\bm{e_{k_n}}]$ with $b_k$ generic scalars and $\bm{e_k}$ the $k$-th canonical vector of dimension $N$ determining the input nodes $\{k_1,k_2,\dots,k_n\}$ of the graph formed by this system. In this paper, without loss of generality, we mark $k_1$ is node $1$. $C_n$ is a diagonal matrix with $c_n$ as its entries. $C_a$ is formed by the corresponding $c_a^+$ and $c_a^-$.

\section{Dynamic Properties of the Model}
We next discuss four essential properties of this dynamic neural network.
\subsection{Boundedness and Stability}
From the view of biology, it is reasonable to assume that the external stimuli and the activation function among neurons are bounded. Then, in our network model, we have 
\begin{equation}
	\label{amp1}
	|u_i|\le u_{max},\  |\phi(\cdot)|\le \phi_{max}.
\end{equation}
The commonly used activation function -- sigmoid function satisfies the boundedness assumption. In this paper, we use the range of the sigmoid function to be $[-1,1]$ such that $\sup{\phi} =1$ and $\phi(0)=0$. From the dynamic equations (\ref{eq:model}), we define $\mathcal{X}=\{\bm{x}\in \mathbb{R}^n:|x_i|\le x_{max}\}$, where $x_{max}=\frac{||B\bm{u}||_{\infty}}{c_n-d^{in}\max\{\overline{a}^+,|\underline{a}^-|\}}$ denotes the maximum membrane potential.

\begin{lemma}
	\label{lemma1}
	When assumptions (\ref{amp1}) hold and $c_n>d^{in}\max\{\overline{a}^+,|\underline{a}^-|\}$, the trajectory of the coupled dynamic system (\ref{eq:model}) from any starting point $\bm{x_0}\in\mathcal{X}$ will remain in $\mathcal{X}$. 
\end{lemma}
\begin{proof}
    The lemma asserts that $\mathcal{X}$ is an invariant set. Consider a solution $\bm{x^s}$ of $(\ref{2a})$ starting in $\mathcal{X}$ and $|x_i|$ is the largest entry of $\bm{x^s}$. We may assume $x_i=x_{max}=\frac{||B\bm{u}||_{\infty}}{c_n-d^{in}\max\{\overline{a}^+,|\underline{a}^-|\}}$. The right-hand side of $(\ref{1a})$ is $-c_nx_i+\gamma_\theta(\sum_{(j,i)} a_{ij}x_j)+b_iu_i$.
    If $-\theta<\sum_{(j,i)} a_{ij}x_j<\theta$, it becomes $-c_nx_i+b_iu_i=\frac{-c_n||B\bm{u}||_{\infty}}{c_n-d^{in}\max\{\overline{a}^+,|\underline{a}^-|\}}+b_iu_i<-||Bu||_{\infty}+b_iu_i\le 0$. If $\sum_{(j,i)} a_{ij}x_j<-\theta$ or $\sum_{(j,i)} a_{ij}x_j>\theta$, it becomes $-c_nx_i+\sum_{(j,i)} a_{ij}x_j+b_iu_i\le (-c_n+\sum_{(j,i)} |a_{ij}|)x_i+b_iu_i\le (-c_n+\max\{d^{in}\overline{a}^+,d^{in}|\underline{a}^-|\})x_i+b_iu_i=-||Bu||_{\infty}+b_iu_i\le 0$. It can be observed that both cases drive  $x_i$ into the interior or in the worst case to a point where $\vert x_i\vert = x_{max}$. It is the same when $x_i$ approaches $-x_{max}$. Therefore, $\mathcal{X}$ is an invariant set. %Another case is when a solution to the equation $(\ref{2a})$ has initial conditions outside $\mathcal{X}$, say $|x_i|>x_{max}=\frac{||Bu||_{\infty}}{c_n-d^{in}\max\{\overline{a}^+,|\underline{a}^-|\}}$, where $|x_i|$ is the largest entry of $\bm{x^s}$. Without loss of generality, we assume $x_i>0$. When $d^{in}\max\{\overline{a}^+,|\underline{a}^-|\}<c_n$, the right-hand side of $(\ref{1a})$ becomes $-c_nx_i+\gamma_\theta(\sum_{(j,i)} a_{ij}x_j)+b_iu_i\le (-c_n+\sum_{(j,i)} |a_{ij}|)x_i+b_iu_i\le (-c_n+\max\{d^{in}\overline{a}^+,d^{in}|\underline{a}^-|\})x_i+b_iu_i<-||B\bm{u}||_{\infty}+b_iu_i\le 0$ until $x_i=x_{max}$. It means the solution will decrease until it evolves into the set $\mathcal{X}$. Therefore, we can conclude that $\mathcal{X}$ is an invariant absorbing set. 
	 %and a similar proof has been given in \cite{menara2019}, so that it is omitted here.
\end{proof}

This boundedness feature keeps the modeled neural connectome and activations within finite ranges consistent with empirical observations. Additionally, the bounded individual neural states reflect how membrane potentials saturate under extreme inputs. 

Biological neurons are not electrically active at rest, with membrane potentials remaining in a steady-state baseline condition. Upon experiencing external stimuli, neural membrane potentials deviate from this baseline due to ion fluxes across the cell membrane induced by the stimulus. When the stimulus subsides, the membrane potential returns to the original baseline steady-state.  Our proposed system (\ref{eq:model}) also exhibits this feature.

\begin{theorem}
	\label{thm1}
	When conditions in lemma \ref{lemma1} are satisfied, the neuromimetic dynamic model (\ref{2a}) is asymptotically stable in the absence of external stimuli.% and the whole system ($\ref{eq:model}$) is stable.
\end{theorem}

Since conditions in Lemma \ref{lemma1} establish that the infinity norm qualifies as a Lyapunov function, we omit the detailed proof of the Theorem \ref{thm1}.

\subsection{Structural Controllability and Observability}
Because brain networks are enablers of cognition and behavior, a growing body of literature is focused on the control theory of brain network models.
Here, our goal is to extend concepts of structural controllability and observability to nonlinear hybrid system models that involve network dynamics coupled with the dynamics of groups of neurons and brain regions. It is noted that most work on structural controllability and observability by \cite{li1996cactus,lin1974structural,mayeda1981structural} focuses on LTI systems, and proof of the (structural) controllability of a hybrid system that we consider next requires dealing with time-varying dynamics, making it somewhat challenging to determine conditions for controllability. %In reality, not all brain regions are engaged and activated if the external stimulus is small. 
In this section, controllability (observability) of such systems is studied using graph theory. From (\ref{2b}), we know that the graph topology of our network models remains static over time, while only the connection strengths change according to hybrid dynamics (\ref{eq:model}). Specifically, the time-varying weight matrix $A$ dictate how neuron activities influence each other. %Therefore, in each time slot, the whole system is a hybrid piecewise affine system. %We draw on the definition of structural controllability for a linear time-varying system. 
\begin{comment}
    
\begin{definition}
	\label{definition1}
	 The control graph $\mathcal{G}$ associated with ($\ref{2a}$) is structurally controllable if there exists a linear controllable system having the same structure as $\mathcal{G}$.
\end{definition}
\end{comment}
%The definition of \emph{structural observability} is given similarly. %that there exists a observable linear time varying system having the same structure with $\mathcal{G}$. 
Next, we provide an analysis of structural controllability and observability of the graph generated by our model. Based on Definition \ref{definition_scg}, it can be inferred that the system (\ref{eq:model}) is structurally controllable (observable) if there exists a controllable (observable) graph with the same structure. %This is due to the fact that across different time slots, the weight matrix $A$ evolves while maintaining its structural integrity. 
Specifically, the locations of zero entries remain constant, with only non-zero entries subject to variation, thereby preserving the underlying graph topology. Consequently, the subsequent analysis will concentrate on a LTI system associated with this graph topology. %It disregards the influence of the zero flat interval in the activation function $\Gamma(\cdot)$, thereby focusing on the core structural properties of the system.

\begin{theorem}
	\label{thm2}
	The digraph $\mathcal{G}$ associated with an LTI system having multiple control input nodes $\{k_1,k_2,\dots,k_n\}$ and output nodes $\{o_1,o_2,\dots,o_n\}$ is structurally controllable and structurally observable if it possesses a generalized cactus structure rooted from the control nodes and terminated at the output nodes.
\end{theorem}

\begin{proof}
	From the above analysis, we know that to prove the digraph $\mathcal{G}$ of the model (\ref{eq:model}) is structurally controllable and structurally observable is same as finding a controllable and observable graph with the corresponding weight matrix $A$, input matrix $B$ and output matrix $C$. Based on the results in \cite[Proposition 3]{lin1974structural} and \cite[Theorem 1]{rech1990structural}, it is concluded that a digraph with a single control node and a single output node is structurally controllable and observable if it is spanned by a cactus rooted at the control node and terminated at the output node. Then, each subgraph $\mathcal{G}_{k}$ comprised of a cactus rooted at input node $k$ with $k\in\{1,2,\dots,n\}$, is structurally controllable, and we can find a choice of weights to form a weight matrix $A_{k}$ and corresponding $B_{k}$ such that $(A_k,B_k)$ is a controllable pair. This corresponds to a generalized cactus structure with disconnected components. Meanwhile, we choose $\underline{A},\overline{A}$ to guarantee $\underline{A}\preceq  A_{k}\preceq \overline{A}$ for all $k$. If we combine all these disjoint subgraphs, the weight matrix can be written as $A'=\begin{bmatrix}
		A_1& \mathbf{0} &\cdots  & \mathbf{0}\\
		\mathbf{0}& A_2 &\mathbf{0}  &\mathbf{0} \\
		\vdots& \vdots &\ddots   &\vdots \\
		\mathbf{0}& \mathbf{0} &\cdots  &A_n
	\end{bmatrix}$ with $B'=[b_{k_1}\bm{e_{k_1}},\ b_{k_2}\bm{e_{k_2}},\ \dots,\ b_{k_n}\bm{e_{k_n}}]$. Given that $(A_{k},B_{k})$ is a controllable pair for all $k$, the controllability matrix $\mathcal{W}_{k}$ is full rank. Therefore, the controllability matrix formed by $(A',B')$ has the form $\begin{bmatrix}
		\mathcal{W}_1& \mathbf{0} &\cdots  & \mathbf{0}\\
		\mathbf{0}& \mathcal{W}_2 &\mathbf{0}  &\mathbf{0} \\
		\vdots& \vdots &\ddots   &\vdots \\
		\mathbf{0}& \mathbf{0} &\cdots  &\mathcal{W}_n
	\end{bmatrix}$ after rearranging columns, which is easily observed to be full rank. 
 
%Suppose $A'$ is perturbed by a small perturbation matrix $\delta$ resulting from adding new edges with tiny weights. By continuity of the rank function, if $\delta$ is sufficiently small, controllability property is preserved.
If more edges are added among subgraphs $\mathcal{G}_k$, it does not introduce inaccessible nodes or otherwise violate structurally controllable, and thus the structural controllability property is preserved. %According to \cite{lin1974structural}[Theorem], we come to the same conclusion that adding more edges of a controllable graph does not influence its structural controllability. 
 To see this, we prove the following \underbar{claim}. If edges between the first two subgraphs $\mathcal{G}_1$ and $\mathcal{G}_2$ are added, the weight and input matrix of these two subgraphs are $A'=\begin{bmatrix}
			A_1& E_{12} \\
			E_{21}& A_2
		\end{bmatrix}$ and $B'=\begin{bmatrix}
			B_1& \mathbf{0} \\
			\mathbf{0}& B_2
		\end{bmatrix}$, where $E_{12}$ and $E_{21}$ represent the added edges. According to Popov-Belevitch-Hautus controllability critieria, if $(A',B')$ is a controllable pair, no $\bm{z}=[\bm{z_1};\bm{z_2}]\neq\bm{0}$ satisfies $\bm{z}^TA'=\omega\bm{z}^T$ and $\bm{z}^TB'=\bm{0}$ simultaneously. 
 Suppose there exist $\bm{z_1}$ and $\bm{z_2}$ not both equal to $\bm{0}$ yet satisfying the above two equations. Then, we have
\begin{subequations}
\label{eq:6}
 \begin{align}
     \bm{z_1}^TA_1+\bm{z_2}E_{21}&=\omega\bm{z_1}^T,\label{6a}\\
     \bm{z_1}^TE_{12}+\bm{z_2}^TA_2&=\omega\bm{z_2}^T,\label{6b}\\
     \bm{z_1}^TB_1=\bm{0},\ \ \bm{z_2}^T&B_2=\bm{0}.\label{6c}
 \end{align}
\end{subequations}
 (a) Assume that $\bm{z_1}=\bm{0}$, then $\bm{z_2}\neq\bm{0}$. Since $(A_1,B_1)$ is a controllable pair, $\bm{z_2}^TA_2=\omega\bm{z_2}^T$ and $\bm{z_2}^TB_2=\bm{0}$ implies $\bm{z_2}=\bm{0}$. According to (\ref{6b}) and (\ref{6c}), $\bm{z_2}=\bm{0}$ when $\bm{z_1}=\bm{0}$, which contradicts the assumption that $z\ne 0$. It is the same situation when $\bm{z_2}=\bm{0}$ and $\bm{z_1}\neq\bm{0}$. Therefore, neither $\bm{z_1}$ nor $\bm{z_2}$ is zero.\\
(b) Assume that $E_{12}$ is a zero matrix. According to (\ref{6b}), (\ref{6c}) and the assumtion that $(A_2,B_2)$ is a controllable pair, $\bm{z_2}=\bm{0}$. Then, (\ref{6a}) becomes $\bm{z_1}^TA_1=\omega\bm{z_1}^T$, which implies $\bm{z_1}=\bm{0}$, which again contradicts our original assumption. It is the same situation when $E_{21}$ is zero. Therefore, neither $E_{12}$ nor $E_{12}$ is zero.\\
(c) Assume that $\bm{z_1}^TE_{12}=\bm{0}$. Equations (\ref{6b}), (\ref{6c}) imply $\bm{z_2}=\bm{0}$ and then from (\ref{6b}), we have $\bm{z_1}=\bm{0}$, which contradicts the assumption that $(A_2,B_2)$ is controllable. It is the same situation when $\bm{z_2}^TE_{21}=\bm{0}$.\\
Till now, we have the conditions $\bm{z_1}$, $\bm{z_2}$, $E_{12}$, $E_{21}$, $\bm{z_1}^TE_{12}$, $\bm{z_2}^TE_{21}\neq\bm{0}$. From (\ref{6a}) and (\ref{6b}), $\omega$ can be any complex number except eigenvalues of $A_1$ or $A_2$, then $det(\omega I-A_1)$, $det(\omega I-A_2)\neq0$. We may multiply $(\omega I-A_1)^{-1}B_1$ and $(\omega I-A_2)^{-1}B_2$ on both sides of equations (\ref{6a}) and (\ref{6b}), respectively. Equations become $\bm{z_1}^TB_1=\bm{z_2}^TE_{21}(\omega I-A_1)^{-1}B_1=\bm{0}$ and $\bm{z_2}^TB_2=\bm{z_1}^TE_{12}(\omega I-A_2)^{-1}B_2=\bm{0}$. If the above two equations always hold for all $\omega$, terms $\bm{z_1}^TE_{12}$ and $\bm{z_2}^TE_{21}$ should be zero, which contradicts the assumption.\\
Therefore, there exists no $\bm{z}\neq0$ that satisfies $\bm{z}^TA'=\omega\bm{z}^T$ and $\bm{z}^TB'=\bm{0}$ simultaneously, so that $(A',B')$ is a controllable pair. When adding edges between more than two subgraphs, the proof steps are the same by considering new subgraph as $A_2$ and the graph obtained in the previous step is $A_1$. 

Thus, according to definition \ref{definition_scg}, $\mathcal{G'}$ is a structurally controllable graph and the corresponding dynamic system is structurally controllable.
	The proof that $\mathcal{G'}$ is structurally observable is completely analogous, and the proof is omitted. 
\end{proof}

\begin{remark}
	(Symmetric Graph) In our model $(\ref{eq:model})$, there is no restriction between $(i,j)$ and $(j,i)$. However, recent work by \cite{galan2008network} proposes a general noise-free Wilson–Cowan 
 model in which our weight matrix $A$ would be assumed to be a symmetric matrix that represents the anatomical connectivity of the brain. Here, we note that that a symmetric weighted digraph formed by the dynamic system (\ref{eq:model}) is structurally controllable with multiple control nodes $\{k_1,k_2,\dots,k_n\}$ provided the digraph has a sym-cactus structure rooted at $\{k_1,k_2,\dots,k_n\}$, as defined in \cite{sun2024neuromimetic}). Proofs follow along lines similar to \cite[Theorem 3.3]{menara2018structural} or can follow steps similar to Theorem \ref{thm2} above.
\end{remark}

\subsection{Resilience and Cascade}
Coordinated movement depends on intentionality, prediction, sensory processing, and resilience in neural pathways. Experiments show animals can adapt to disrupted pathways by forming new ones or using alternative intact connections. This demonstrates built-in adaptability and redundancy in sensory-motor neural networks in the brain, enabling continued activity despite damage. The brain network maintains function by utilizing alternative routes when existing ones are disrupted. Our model (\ref{eq:model}) exhibits this characteristic as well. When considering that connections among neurons may be intermittently unavailable, we have the following lemma.
\begin{comment}
the model becomes
\begin{subequations}
	\label{eq:resilience_model}
	\begin{align}
		\dot x(t)&=-C_nx(t)+\Gamma(P_c\circ A(t)x(t))+ Bu,\,\label{5a}\\
		%\dot A&=-C_aA+\phi(xx^T),
		A(t+\tau)&=\left[C_a\circ A(t)+\phi(x(t)x(t)^T)\right]_{\underline A}^{\overline A},\label{5b}\\
		y(t)&=Cx(t),\label{5c}
	\end{align}
\end{subequations}
where $P_c$ is a $n\times n$ matrix with all entries equal $1$ except the resilence connections $(i,j)$ from node $j$ to $i$ are zero.
\end{comment}
\begin{lemma}
	\label{lemma3}
	Suppose there exists a set of neurons $\mathcal{S}\ \subset\  \mathcal{V}$ and connections $\mathcal{E'}\subset\mathcal{E}$ that are intermittently available. If the connectivity graph $\mathcal{G}$ and its subgraph $(\mathcal{V}-\mathcal{S},\mathcal{E}-\mathcal{E'})$ both have a generalized cactus architecture related to the same input nodes,  then the dynamic system (\ref{eq:model}) remains stable, structurally controllable and observable.
\end{lemma}
\begin{proof}
	It is obtained graphically. From Theorem \ref{thm2}, we obtain that as long as the graph preserves the generalized cactus structure during dynamic evolutions, the system is structurally controllable and observable relative to the same input nodes. Therefore, even if there are neurons and connections that are intermittently available, the system is always structurally controllable and observable. Meanwhile, even if there are edges and nodes unavailable, conditions in lemma \ref{lemma1} still hold. Thus, the state matrix in (\ref{2a}) is still a diagonally dominant matrix and thus, the system remains stable.
\end{proof}
\begin{remark}
    (Growth and Pruning of Connections) The current model focuses on synaptic plasticity within a fixed network topology, without accounting for the formation of novel connections or the elimination of lesser-utilized synapses. However, it is essential to incorporate the dynamics of connection growth and pruning to accurately describe the processes underlying developmental neurobiology. According to Theorem \ref{thm2}, the system retains its structural controllability despite the removal or addition of connections, contingent upon the preservation of the generalized cactus topology.
\end{remark}

\begin{figure}[h]
	\begin{center}
		\includegraphics[scale=0.263]{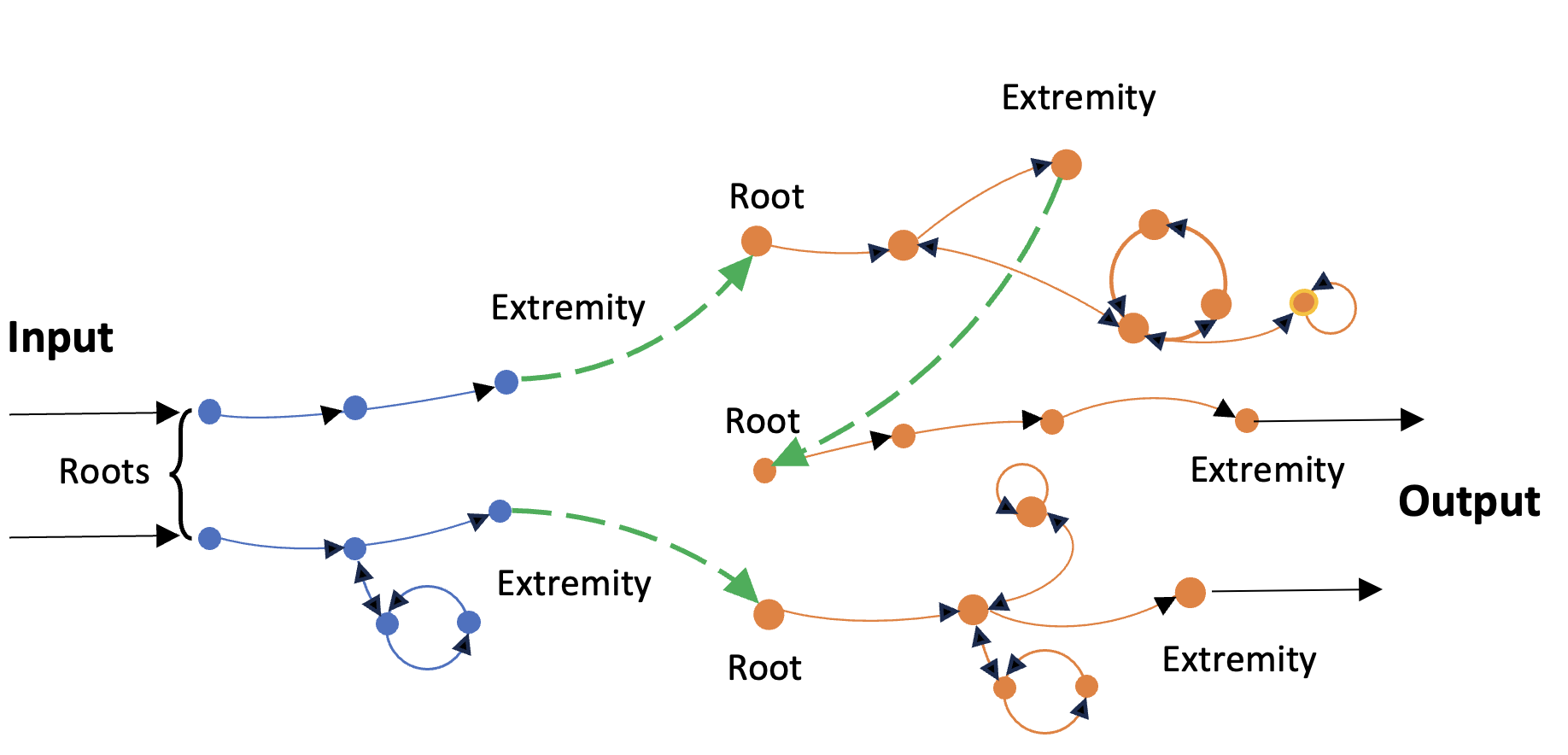}
	\end{center}
	\caption{ A cascaded-cactus structure with 5 subsystems and two of them (blue) have external inputs. Green lines are the interconnections between two subsystems.}
	\label{fig:cascade}
\end{figure}

Meanwhile, the model can be generalized to accommodate \emph{cascaded subsystems}. A related work on composite linear time-invariant systems is \cite{carvalho2017composability}, which focuses on deriving conditions for the structure of the subsystems and their interconnections to ensure that the serial systems are structurally controllable. In contrast, our approach relies on cascading structurally controllable subsystems or extracting a subgraph from all subsystems to render the new system structurally controllable. This method circumvents the need for deriving explicit conditions on the subsystem structures and their interconnections, instead leveraging the cactus topology of the individual subsystems or their subgraphs to establish structural controllability for the composite system.
\begin{definition}(Cactus subgraph)
\label{definition_cactus-sungraph}
    Given a control graph with a generalized cactus structure, a cactus subgraph is a set of connected nodes and edges such that one could obtain a cactus graph — as defined by \cite{lin1974structural} — by identifying a hypothetical root node which would make the subgraph a cactus control graph in its own right if a control input was attached to the identified root node.
\end{definition}
\begin{definition}
	\label{cascaded}
	(Cascaded-cactus Graph)
	%A graph with multiple input nodes is cascaded if it satisfies: for any subgraph, its root node either connects to the extremity of another subgraph or directly receives the external input. Additionally, each extremity links at most one subgraph.
 A graph with multiple input nodes is said to be cascaded if it satisfies the following conditions: for any cactus subgraph, its root node either connects to the terminal node of another subgraph or receives direct external input. Additionally, each terminal node of the cactus subgraph $\mathcal{G}_i$ can connect to at most one root of another cactus subgraph $\mathcal{G}_j$, unless $\mathcal{G}_j$ only contains a cycle.
	
\end{definition}
%This definition naturally generalizes to multi-input scenarios whereby each input line manifests as a  cascading cactus architecture. 
Fig. \ref{fig:cascade} gives an example of five subsystems constructing a cascade-cactus structure. We note that the representation of a generalized cactus structure as a cascade of generalized cactus graphs is generally not unique.

%Consider a set of single-input structurally controllable subsystems $\mathcal{P}=\{S_1,S_2,...,S_P\}$, where only a strict subset have specified external inputs. Without loss of generality, we index the subsystems so that the first $U$ members, i.e., $\mathcal{U}=\{S_1,S_2,...,S_U\}$ with $U<P$ of the set $\mathcal{P}$ constitute those receiving external inputs.

\begin{comment}
	\begin{lemma}
		If the root nodes of subsystems $\mathcal{P}/\mathcal{U}$ connect to extremities of $\mathcal{U}$, with any such extremity linking at most one non-input subsystem, then the whole system with the cascaded graph and its spanning graphs exhibit structural controllability and observability.
	\end{lemma}
	However, in typical scenarios with $U \ll P$, the number of input-receiving subsystems constitutes a small fraction. Consequently, not all non-input subsystems can be matched to an input-receiving subsystem in $\mathcal{U}$.
\end{comment}

\begin{theorem}
	\label{thm:cascade}
	Given a system possessing a (spanned) cascaded-cactus structure with multiple input nodes is structurally controllable and structurally observable. If each its subgraph evolves according to the dynamics describe in $(\ref{eq:model})$, the system is stable and resilient to neuron and connection dropouts when the condition in Lemma \ref{lemma3} holds. 
\end{theorem}
\begin{proof}
	The formation of a cascaded-cactus structure via introducing additional connections between subsystems graphically corresponds to prolonging the stem of the prior subsystem. Thus, the resultant graph preserves a generalized cactus architecture conditional on the base subsystems exhibiting individual cactus topologies. According to Theorem \ref{thm2}, we obtain that the cascaded-cactus configuration is structurally controllable and observable. Concurrently, if all nodes obey the dynamical equation (\ref{eq:model}), the maximum in-degree $d^{in}$ is unchanged. Consequently, the stability conditions given in Lemma \ref{lemma1} hold with the bounds stipulated in Theorem \ref{thm1}. Thereafter, Lemma \ref{lemma3} directly implies resilience of the system to both neuron ($x$) and connection ($a_{ij}$) dropouts.
\end{proof}
\begin{remark}
In complex biological systems, such as animal brain regions, there are often numerous external inputs, contrasting with simpler models that only receive inputs from leftmost root nodes (as depicted in Figure \ref{fig:cascade}). In reality, each subsystem within such a complex structure may possess individual inputs, allowing it to evolve independently and manifest diverse functionalities. However, in scenarios where external inputs are limited, a cascaded structure plays a crucial role. This structure ensures the structural controllability of the entire system, even though certain specific features may be influenced. It is important to note that this structural controllability comes at a cost: the control energy typically increases under these circumstances.
This phenomenon highlights the trade-off between system complexity, input distribution, and control efficiency in biological and other complex networks. 
\end{remark}

Evidence presented in the studies by \cite{agosta2013disruption} on primary progressive aphasia demonstrates disrupted structural connectivity in both the dorsal and ventral speech pathways in these patients. Utilizing MRI techniques, they concluded that damage to white matter connections between cortical areas can result in neural dysfunction. Current theories of brain structure propose that cognitive functions comprise distributed, segregated, yet overlapping networks. Thus, while certain white matter disconnections may lead to specific cognitive impairments, other functional networks remain intact. As characterized by our model (\ref{eq:model}), structural controllability can be guaranteed as long as the generalized cactus structure is maintained. 

\begin{comment}
    
According to Theorem \ref{thm:cascade}, we also design an Algorithm \ref{alg} that can construct a multi-input cascade-cactus that is stable, structurally controllable and observable.
\begin{algorithm2e}
\caption{Constructing a structurally controllable multi-input cascade-cactus system}\label{alg}
\State Input: The set of all subsystems $\mathcal{P}$ and the set of subsystems with external inputs $\mathcal{U}$;\\
 \While{$\mathcal{P}-\mathcal{U}$ is not empty}{
        \State Choose any subsystem $S_u$ belongs to $\mathcal{U}$ and a subsystem $S_{p-u}$ belongs to $\mathcal{P}-\mathcal{U}$;
	\State Make a unidirectional connection between the extremity of $S_u$ and the root of $S_{p-u}$ and denote this generated subsystem $S_{u'}$;
	\State Replace $S_u$ with $S_{u'}$ in the set $\mathcal{P}$ and $\mathcal{U}$. Delete the subsystem $S_{p-u}$ in $\mathcal{P}$;
 }
 \State Output: The disjoint union of all elements in $\mathcal{U}$ constitute a cascade-cactus system, where their roots receiving external inputs and their extremities are the output nodes.
\end{algorithm2e}

However, it is challenging to determine whether a preceding subsystem is structurally controllable or not in some cases. Consequently, the structural controllability of the cascaded graph is difficult to verify. The following study investigates this situation. 
\end{comment}

If two control system graphs are composed in a cascade as described in Definition \ref{cascaded}, it is of obvious interest to know whether structural controllability of the components is either necessary or sufficient for the composite system to be structurally controllable.  It turns out that structural controllability of the components is not necessary in order for the cascade of the two systems to be structurally controllable.

\begin{theorem}
\label{thm:cascade1}
(Condition for structural controllability a single input system driven by the output of another system) Consider two dynamic systems with graph topologies $\mathcal{G}_1$ and $\mathcal{G}_2$. Assume that $\mathcal{G}_2$ is structurally controllable, but the controllability of $\mathcal{G}_1$ is unknown. Let the input nodes and output nodes of $\mathcal{G}_i$ be denoted as ${k_i^1, k_i^2, \cdots}$ and ${o_i^1, o_i^2, \cdots}$, respectively, where $i=1,2$.
Assume that the graph $\mathcal{G}_2$ only has a single input node. If there exists at least one path starting from any input node of $\mathcal{G}_1$ and ending at the input node $k_2^1$ of $\mathcal{G}_2$, then the set of all nodes on this path, together with the nodes in $\mathcal{G}_2$, form a new structurally controllable graph.
\end{theorem}
\begin{proof}
$\Rightarrow:$ Assume that there are a total of $k$ nodes on the path from the input node $k_1^1$ to the end $k_2^1$ (not including $k_2^1$). All these nodes form a connectivity matrix $A_1$ with edge weight, say $1$, and input matrix $\bm{b}_1$ that may be written as
\begin{equation}
A_1=\begin{bmatrix}
  0& 0 &\cdots  &0 &0\\
  1& 0 & \cdots &0 &0\\
  0&  1& \cdots &0 &0\\
\vdots &  \vdots& \ddots  &\vdots&\vdots \\
  0&  0& \cdots &1&0
\end{bmatrix}, \bm{b}_1=\begin{bmatrix}
 1\\
 0\\
 0\\
 \vdots\\
 0
\end{bmatrix}. 
\end{equation}
From the form of $A_1$ matrix, the controllability matrix $[\bm{b}_1,A_1\bm{b}_1,A_1^2A\bm{b}_1,\cdots,A_1^{k-1}\bm{b}_1]=\bm{I}_k$, which is full rank. Therefore, the subsystem $\mathcal{G}_1'$ formed by the nodes on the path is controllable even if the whole system $\mathcal{G}_1$ may be uncontrollable. The cascaded system of $\mathcal{G}_1'$ and $\mathcal{G}_2$ gives the connectivity matrix $A=\begin{bmatrix}
  A_1& \bm{0} \\
  \begin{matrix}
  \bm{0}&1 \\
  \bm{0}&\bm{0}
\end{matrix} & A_2 
\end{bmatrix}$ and input matrix $\bm{b}=\begin{bmatrix}
 \bm{b}_1\\
 \bm{0}
\end{bmatrix}$, where $A_2$ is the connectivity matrix of $\mathcal{G}_2$. 
Next, we prove $(A,\bm{b})$ is a structurally controllable pair, which is equivalent to prove that there exists no $\bm{z}\neq 0$ such that $\bm{z}^T\hat A=\omega\bm{z}^T$ and $\bm{z}^Tb=0$, where $\hat A$ has the same structure as $A$. Without loss of generality, we can assume all eigenvalues of $\hat A$ are distinct. Suppose $(A,\bm{b})$ is not structurally controllable. Then there exists $\bm{z}=(\bm{z_1};
\bm{z_2})\neq\bm{0}$, such that $(\bm{z_1}^T\ \bm{z_2}^T)\begin{bmatrix}
  \hat A_1& \bm{0} \\
  \begin{matrix}
  \bm{0}&1 \\
  \bm{0}&\bm{0}
\end{matrix} & \hat A_2 
\end{bmatrix}=\omega(\bm{z_1}^T\ \bm{z_2}^T)$ and $(\bm{z_1}^T\ \bm{z_2}^T)\begin{bmatrix}
 \bm{b}_1\\
 \bm{0}
\end{bmatrix}=\bm{0}$, which may be written as
\begin{equation}
\label{a}
\begin{split}
&\bm{z_1}^T\hat A_1 + \bm{z_2}^T\begin{bmatrix}
  \bm{0}&1 \\
  \bm{0}&\bm{0}
\end{bmatrix}=\omega \bm{z_1}^T\\
&\bm{z_2}^T\hat A_2=\omega \bm{z_2}^T\\
&\bm{z_1}^T\bm{b_1}=0 \\
&\bm{z_1}, \bm{z_2}\  \text{are not both zero}.
\end{split}
\end{equation}
(a) If $\bm{z_2}^T\begin{bmatrix}
  \bm{0}&1 \\
  \bm{0}&\bm{0}
\end{bmatrix}=\bm{0}$, then $\bm{z_2}^T\bm{b_2}=\bm{0}$, where $\bm{b_2}=\begin{bmatrix}
  1 \\
  \bm{0}
\end{bmatrix}$. Meanwhile, $(\hat A_2,\bm{b_2})$ is a controllable pair and $\bm{z_2}^T\hat A_2=\omega\bm{z_2}^T$, which imply $\bm{z_2}=\bm{0}$. According to (\ref{a}), we have $\bm{z_1}^T\hat A_1=\omega \bm{z_1}^T$ and $\bm{z_1}^T\bm{b_1}=0$. Because $(\hat A_1,\bm{b_1})$ is a controllable pair, this means $\bm{z_1}=\bm{0}$, which contradicts the condition that $\bm{z_1}, \bm{z_2}$ are not both zero.\\
(b) Suppose $\bm{z_2}^T\begin{bmatrix}
  \bm{0}&1 \\
  \bm{0}&\bm{0}
\end{bmatrix}\neq\bm{0}$ and according to (\ref{a}), 
\begin{equation}
\label{b}
\bm{z_1}^T(\omega I-\hat A_1)=\bm{z_2}^T\begin{bmatrix}
  \bm{0}&1 \\
  \bm{0}&\bm{0}
\end{bmatrix}.
\end{equation}
If det$(\omega I-\hat A_1)=0$, $\omega$ is an eigenvalue of $\hat A_1$. Since $\bm{z_2}^T\hat A_2=\omega \bm{z_2}^T$, $\omega$ is also an eigenvalue of $\hat A_2$, which is a contradiction of the assumption. Therefore, det$(\omega I-\hat A_1)\neq 0$. Multiplying $(\omega I-\hat A_1)^{-1}\bm{b_1}$ on both sides of equation (\ref{b}), we have 
\begin{equation}
\label{c}
\bm{z_1}^T\bm{b_1}=\bm{z_2}^T\begin{bmatrix}
  \bm{0}&1 \\
  \bm{0}&\bm{0}
\end{bmatrix}(\omega I-\hat A_1)^{-1}\bm{b_1}=\bm{0}.
\end{equation}
Since $\omega$ is the eigenvalue of $\hat A_2$ which can take an arbitrary value by modifying the non-fixed entries of $A_2$ and $\bm{b_1}\neq0$, the only possibility for equation (\ref{c}) to always hold is $\bm{z_2}^T\begin{bmatrix}
  \bm{0}&1 \\
  \bm{0}&\bm{0}
\end{bmatrix}=\bm{0}$, which contradicts the assumption.\\
In conclusion, conditions (\ref{a}) cannot be satisfied simultaneously, which implies that $(A,\bm{b})$ is a structurally controllable pair and $\mathcal{G}_1'\cup \mathcal{G}_2$ is structurally controllable.\\

$\Leftarrow:$ If $\mathcal{G}_1' \subset \mathcal{G}_1$ and $\mathcal{G}_1'\cup \mathcal{G}_2$ is a structurally controllable graph, then $\mathcal{G}_1'\cup \mathcal{G}_2$ has a generalized cactus structure. Therefore, all nodes are accessible, which implies there exists at least a path from any input node of $\mathcal{G}_1'$ to the starting node $n^2_1$ of $\mathcal{G}_2$.
\end{proof}

\begin{proposition}
\label{thm:cascade2}
    (A sufficient but not necessary condition for the multi-input case) Assume that the graph $\mathcal{G}_2$ has $r$ input nodes. If there exist $r$ distinct paths, each starting from an input node $k_1^j$ of $\mathcal{G}_1$ and ending at a corresponding input node $k_2^j$ of $\mathcal{G}_2$, $j=1,2,\cdots,r$, such that these $r$ paths have no common nodes, then the set of all nodes on these $r$ paths, together with the nodes in $\mathcal{G}_2$, form a structurally controllable graph. 
\end{proposition}
\begin{proof}
$\Rightarrow:$ The structural controllability of the composite system can be easily obtained from the proof of Theorem $\ref{thm:cascade1}$ by adding one path from an input node of $\mathcal{G}_1$ to one input node of $G_2$ in a stepwise fashion until all input nodes  $k_2^j$, $j=1,2,\cdots,r$ have paths from distinct $k_1^j$, $j=1,2,\cdots,r$. The structural controllability property is preserved at each step of this process if these $r$ paths share no nodes in common.\\
$\nLeftarrow:$ The following counterexample shows that the condition is not necessary. Suppose the connectivity matrix of $\mathcal{G}_1$ is $A_1=\begin{bmatrix}
  0&0 \\
  1&0
\end{bmatrix}$, and it has one input nodes $k_1^1$ driving the input matrix $b_1=\begin{bmatrix}
  1 \\
  0
\end{bmatrix}$. The connectivity matrix of $\mathcal{G}_2$ is given by $A_2=\begin{bmatrix}
  0&0&0 \\
  0&0&1\\
  0&1&0
\end{bmatrix}$, and it has two input nodes $k_2^1,k_2^2$ to make $\mathcal{G}_2$ controllable. Assume that they have paths from the input node $k_1^1$ and connecting directly to the same node on this path and the path topology is supposed to be $\begin{bmatrix}
  0&1 \\
  0&1\\
  0&0
\end{bmatrix}$. Then, the path together with $\mathcal{G}_2$ form a new graph whose state and input matrices are given by
\[A=\begin{bmatrix}
  A_1& \bm{0} \\
  \begin{matrix}
  0&1 \\
  0&1\\
  0&0
\end{matrix} & A_2 
\end{bmatrix},\bm{b}=\begin{bmatrix}
  b_1 \\
  \bm{0} 
\end{bmatrix}.\] It is easily checked that the controllability matrix of this new graph has full rank so that it is controllable.

Therefore, the statement in Proposition \ref{thm:cascade2} is a sufficient but not necessary condition.
\end{proof}

\section{Simulations}
In this section, we provide simulations to validate the theory presented in the previous sections. We consider two examples, the second of which is a stylized model of the neuronal network of a primate visual cortex.

\begin{figure}[h]
	\begin{center}
		\includegraphics[scale=0.38]{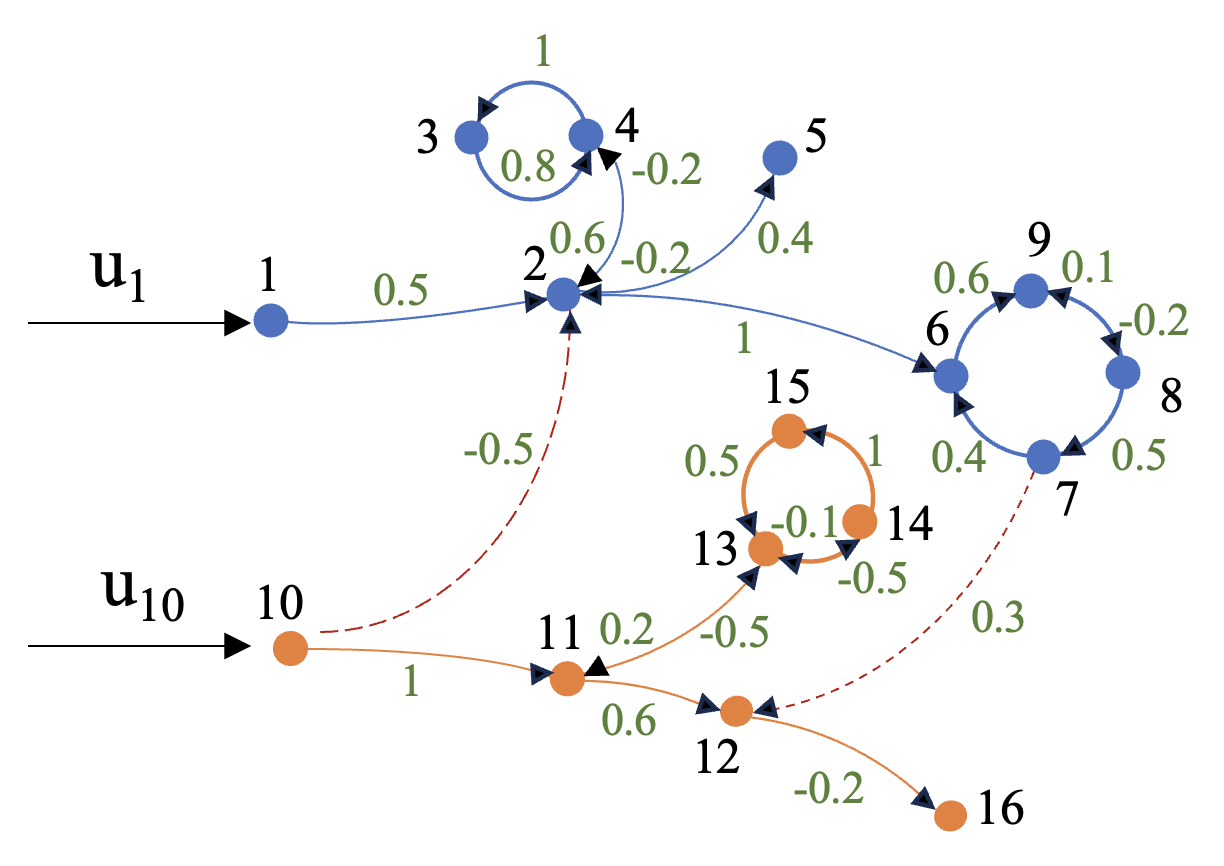}
	\end{center}
	\caption{ A generalized cactus structure with 16 nodes.
		Green numbers denote edge weights at initial conditions. Only nodes 1 and 10 have external input.}
	\label{fig:weight}
\end{figure}

The first example implements the dynamical model (\ref{eq:model}) using a 16-neuron network structured as a generalized cactus graph with two stems, as depicted in Fig. \ref{fig:weight}. This figure shows the weighted directed connections comprising the initialized weight matrix $A$. The maximum in-degree is 4. We simulate the dynamics on this architecture to analyze the control properties of interest. There are two control nodes $k_1=1$ and $k_2=10$, so that $B=[\bm{e_1},\bm{e_{10}}]$ with $b_1=b_{10}=1$ and $\bm{e_1},\bm{e_{10}}\in\mathbb{R}^{16\times 1}$. Meanwhile, we set $\underline{a}^-=-1.2,\overline{a}^-=-0.1$ and $\underline{a}^+=0.1,\ \overline{a}^+=1.2$. $c_a=0.98$ and $\theta=0.1$. The decay rate of each neuron is set to be $-c_n=-5$, which satisfies the condition $c_n-\max d^{in}\{\overline{a}^+,|\underline{a}^-|\}>0$. It is worth noting that $(A,B)$ pair is controllable. The time interval of updating $A$ is $\tau=0.2$s. 

\begin{figure}[h]
	\begin{center}
		\includegraphics[scale=0.225]{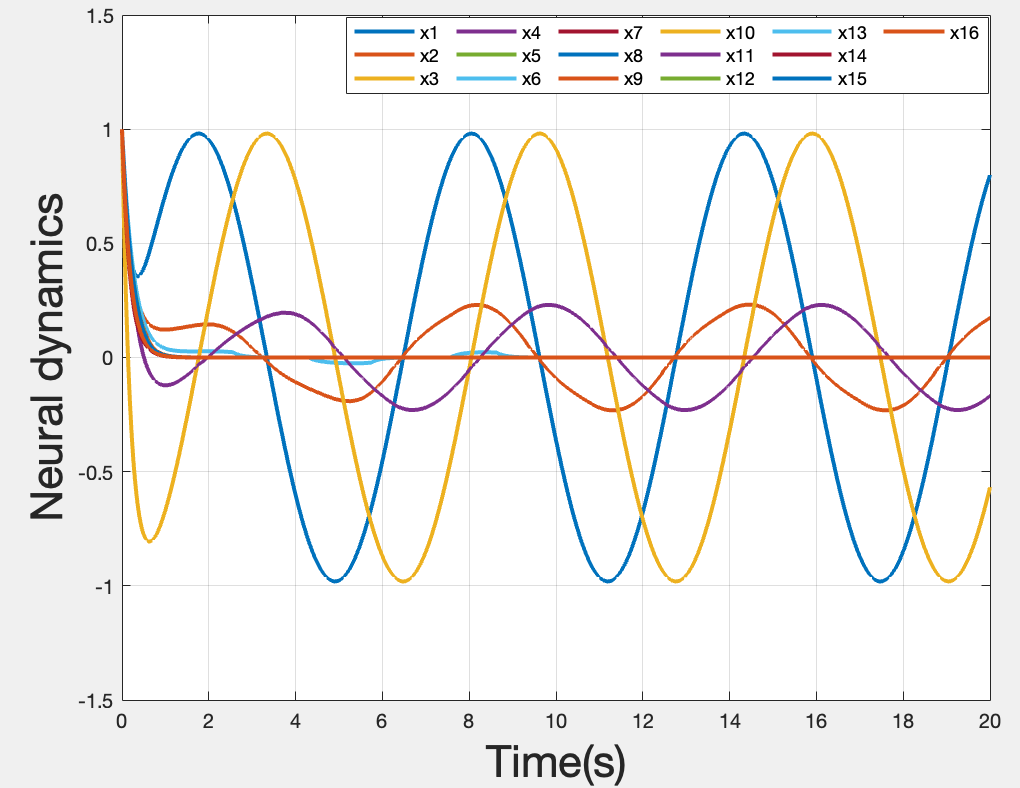}
		\includegraphics[scale=0.225]{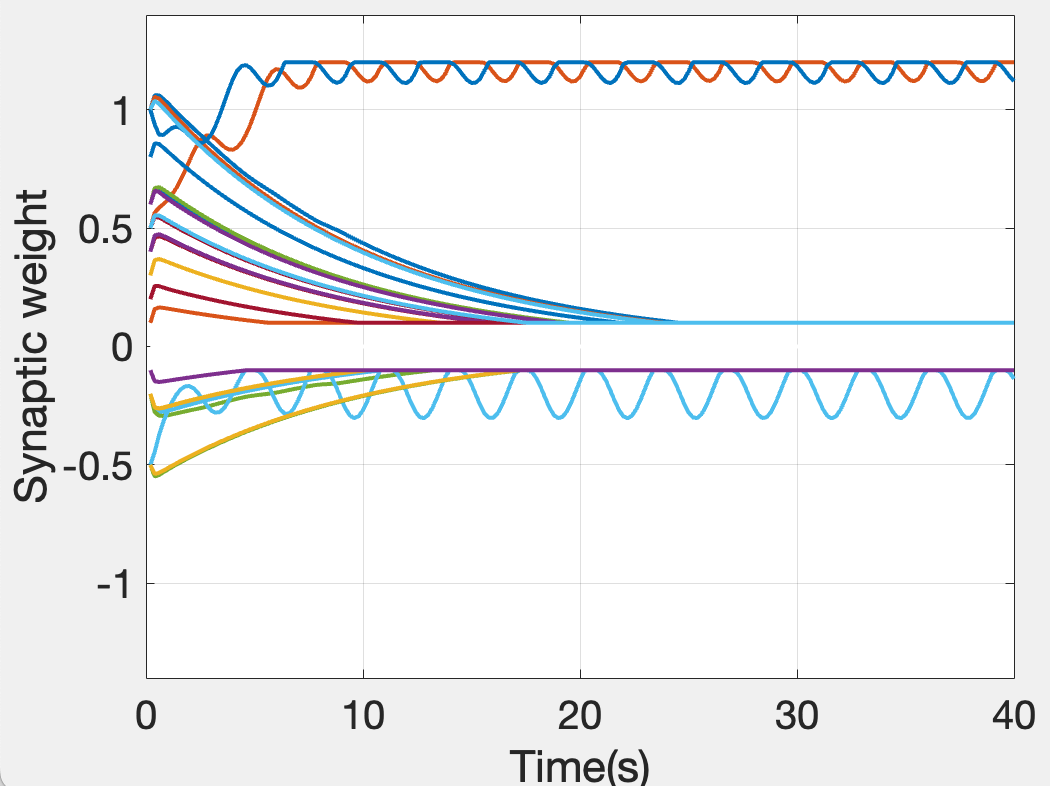}
	\end{center}
	\caption{Figure on the left is an example of the network in Fig. \ref{fig:weight} with external input on node 1 and 10 using the input to be $5\sin(t)$ and $-5\cos(t)$. The right figure depicts the evolution of synaptic weights in 40 seconds.}
	\label{fig:state}
\end{figure}

Suppose all neural states initialize at 1. We set the external inputs to node 1 and node 10 as $u_1=5\sin(t)$ and $u_{10}=-5\cos(t)$, respectively. The results are depicted in Fig. \ref{fig:state}, demonstrating bounded neural dynamics and synaptic weights within the specified ranges identified in Lemma \ref{lemma1}. Overall, these simulations validate the theoretical results on weight boundedness and system stability.

\begin{figure}[h]
	\begin{center}
		\includegraphics[scale=0.33]{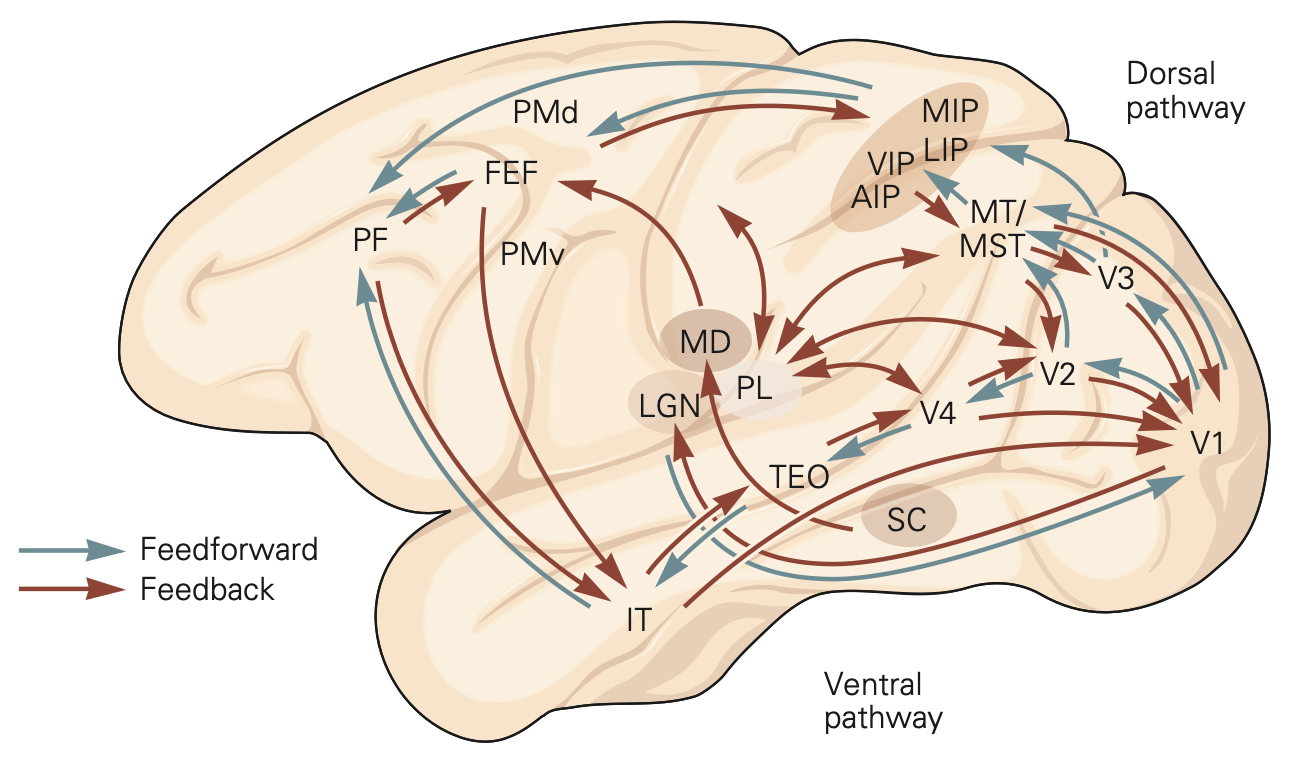}
	\end{center}
	\caption{
The figure retrieved from \cite{2021Pons} illustrates the dorsal and ventral visual pathways involved in action processing. The dorsal pathway projects from the visual cortex to the posterior parietal and frontal cortices. In contrast, the ventral pathway projects to the temporal cortex. The abbreviations for the various brain regions are defined in \cite{2021Pons}.}

	\label{fig:pathway}
\end{figure}

We turn next to a network model of the macaque monkey vision system. Visual processing initiates in the retinas, which extract local features like contrasts and color. Outgoing retinal pathways split into three streams: one through the lateral geniculate nucleus (LGN) to the primary visual cortex (V1), one to the midbrain's pretectal area, and one to the superior colliculus (SC). This divergent architecture enables parallel extraction of distinct visual cues for perception, reflexes, and gaze control. Fig. \ref{fig:pathway} exhibits the visual system of the macaque monkey containing two prominent pathways for processing visual information. The dorsal pathway transmits signals from the primary visual cortex (V1) to the posterior parietal cortex. This pathway is specialized for determining the spatial location and motion of visual stimuli - the "where" properties. Meanwhile, the ventral pathway projects from V1 to the inferior temporal cortex and is optimized for identifying object qualities like shape, color, and category - the "what" properties. 

\begin{figure}[h]
	\begin{center}
		\includegraphics[scale=0.38]{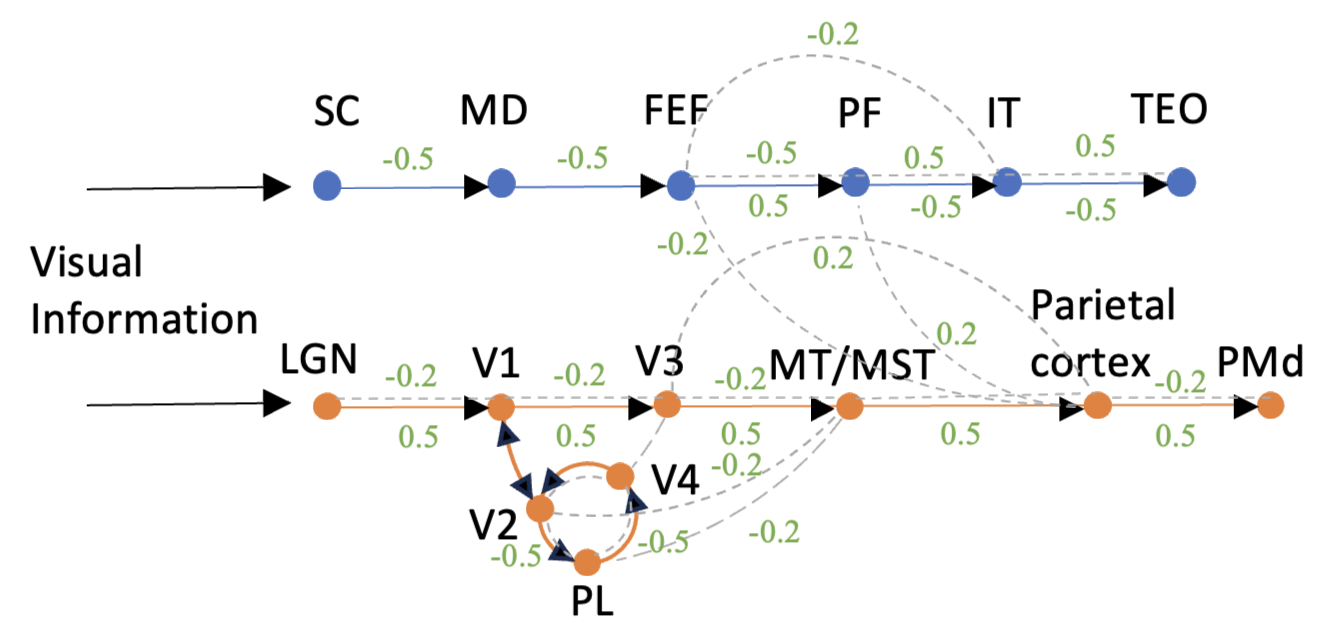}
	\end{center}
	\caption{An abbreviated diagram for partitioning visual pathways to a generalized cactus structure with two stems (in blue and orange) and a bud (in red). Grey dot lines are the remaining connections.}
	\label{fig:diagram}
\end{figure}

The visual processing pathways in Fig. \ref{fig:pathway} can be represented as a generalized cactus graph comprising two primary stems: one originating from the superior colliculus (SC), traversing the mediodorsal nucleus (MD), frontal eye fields (FEF), prefrontal cortex (PF), inferior temporal cortex (IT), and terminating in posterior IT (TEO). Another starts from LGN, passes through V1, V3, MT/MST, and parietal cortex regions (AIP, VIP, LIP, MIP), ending in premotor dorsal cortex (PMd). The bud consists of bidirectional connections between primary visual areas V1, V2, V4 and the pulvinar nucleus (PL), with V1 serving as the joint. An abbreviated diagram is shown in Fig.\ref{fig:diagram}. 

This illustrates one possible scheme for parcellating the visual system into a generalized cactus architecture. Different partitions may enable specialized functional pathways for performing distinct visual processing tasks.

\begin{figure}[h]
	\begin{center}
		\includegraphics[scale=0.236]{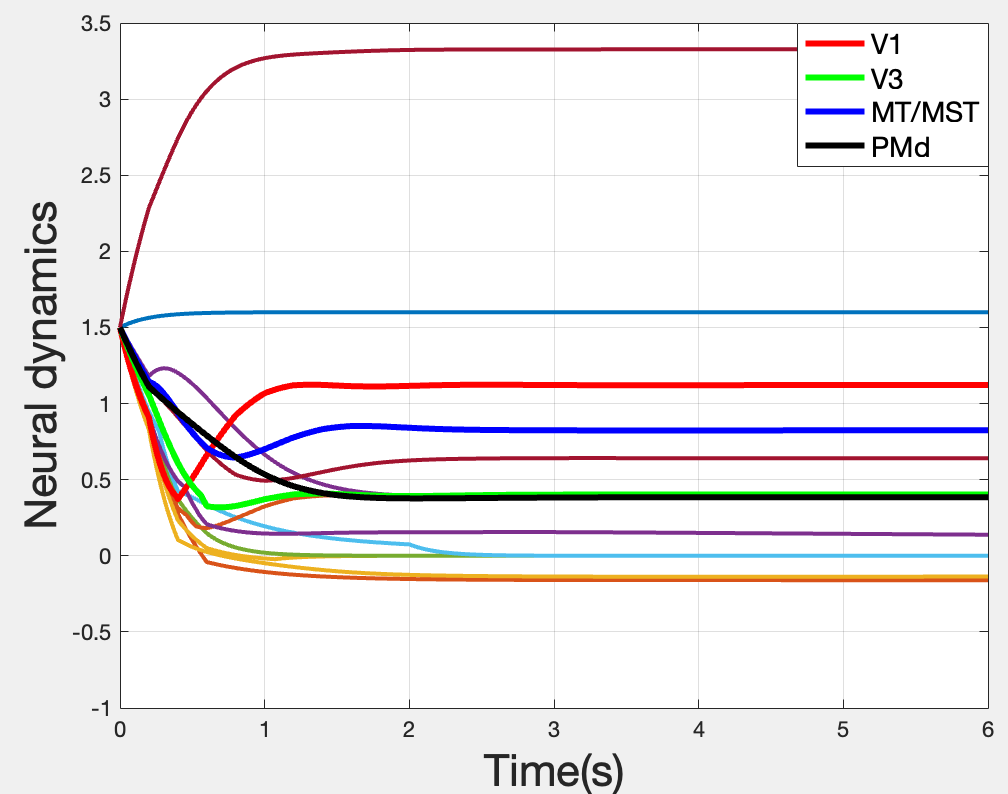}
		\includegraphics[scale=0.236]{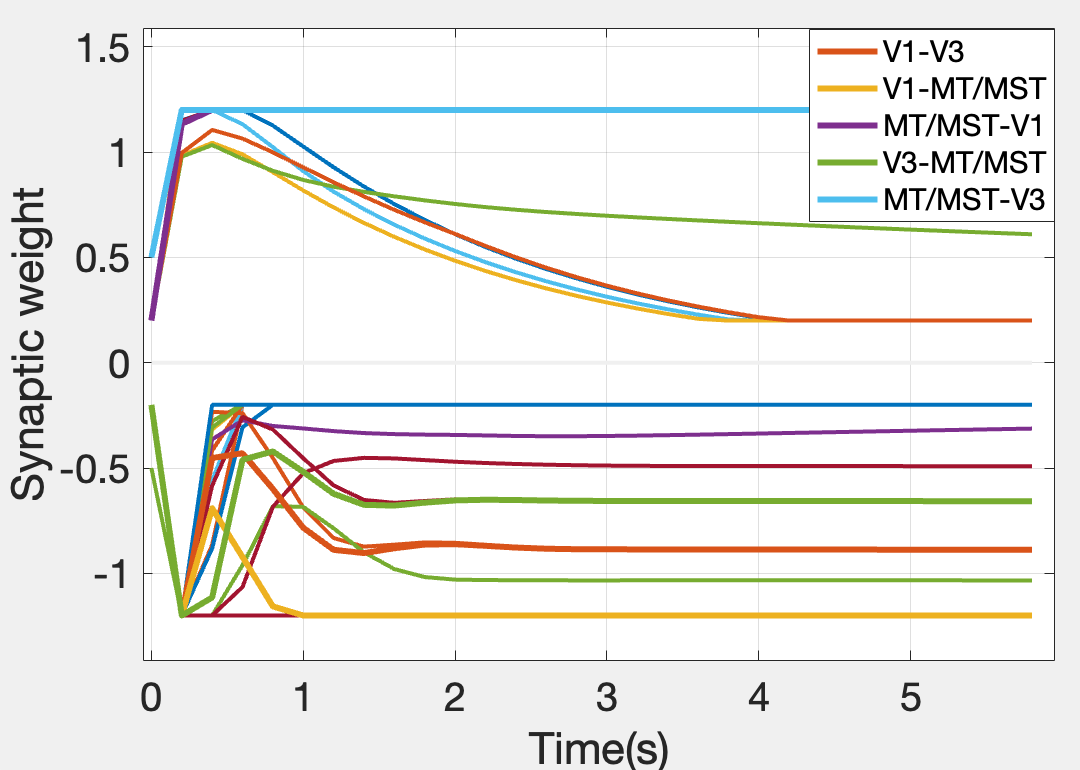}
	\end{center}
	\caption{(a) Simulation of the macaque monkey network with constant control inputs. (b) Synaptic weight dynamics evolving over 6 seconds.}
	\label{fig:constant}
\end{figure}

Linear models such as proposed by \cite{muldoon2016stimulation} and \cite{kim2020linear} have proven useful in elucidating macroscale brain networks work dynamics, but at the microscale neural dynamics are both complex and nonlinear.
 Our proposed model modifies the standard linear brain networks by incorporating nonlinear neuronal properties that have been experimentally observed. However, the connectivity matrices derived from diffusion MRI techniques are symmetric, thus making it difficult to interpret such data in terms of our directed network models. Empirical evidence demonstrates that in the absence of stimuli, neurons exhibit autonomous recovery to baseline resting potential levels. This reflects an inherent stability in neural systems at both the macroscopic network scale and the microscopic single-neuron scale. Here, we appeal to Fig. \ref{fig:diagram} to conduct numerical validation. Suppose external inputs to the SC and LGN are both unit constant values, representing a continuous stimulus from the environment. The initial signs of the synaptic weights are designated as positive (+) for excitatory plasticity and negative (-) for inhibitory plasticity, which corresponds to the feedforward and feedback connections shown in Fig. \ref{fig:pathway}, respectively. For solid line connections, the weights are initialized to 0.5 and -0.5, respectively. Conversely, for dotted line connections, the weights are initialized to 0.2 and -0.2, respectively. The response of different areas in the visual pathways and the synaptic weight dynamics are shown in  Fig. \ref{fig:constant}. We note that the controllability matrices during this evolution remain full rank. 

\begin{figure}[h]
	\begin{center}
		\includegraphics[scale=0.35]{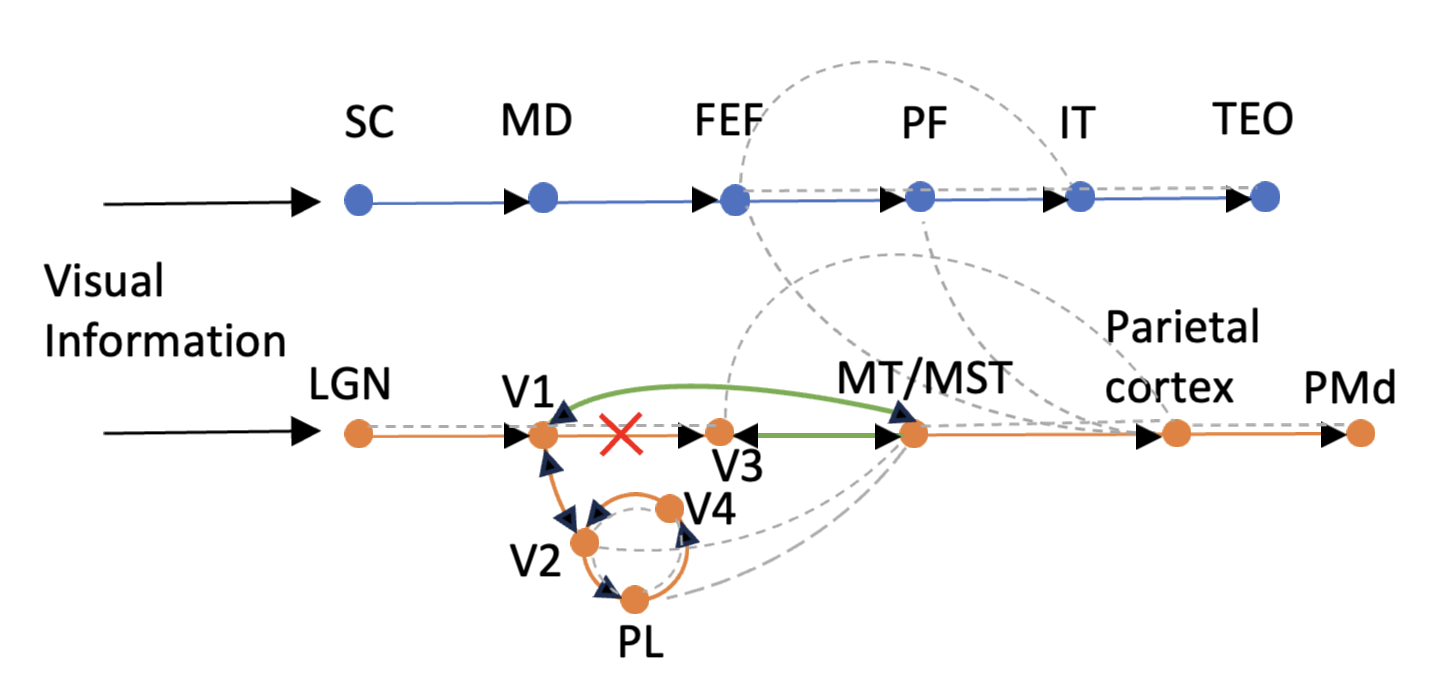}
	\end{center}
	\caption{The diagram of visual pathways without connections between V1 and V3, where green lines exhibit a new bud formed by MT/MST and V3 is taken into account as a backbone.}
	\label{fig:diagram2}
\end{figure}

Next, we simulate what happens if the connections between V1 and V3 are unavailable as might occur due to a brain injury. This renders the network with a new generalized cactus structure as shown in Fig. \ref{fig:diagram2}. We perform the same input to the new graph and note that the controllability matrices during this evolution still remain full rank, which validates that the network maintains structural controllability even with edge deletions, as long as the generalized cactus structure is preserved. This demonstrates the resilience of the controllability property as stated in Lemma \ref{lemma3}. The visual pathways depicted in Fig. \ref{fig:pathway} represent important interconnections that support graphically distinct but functionally equivalent subnetworks within the visual system. Additional routes link these brain regions through alternative channels. Here we particularly consider connections subserving analogous functionality, for example, green lines providing comparable V1 to V3 communication (see Fig. \ref{fig:diagram2}). Generally such redundant connectivity engenders resilience to disruptions caused by brain injury. Our model reflects the fact that neurologic functions in higher animals tend to be resilient in that they can survive the moderate loss of function due to brain injury or disease like stroke.

\begin{figure}[h]
	\begin{center}
            \includegraphics[scale=0.23]{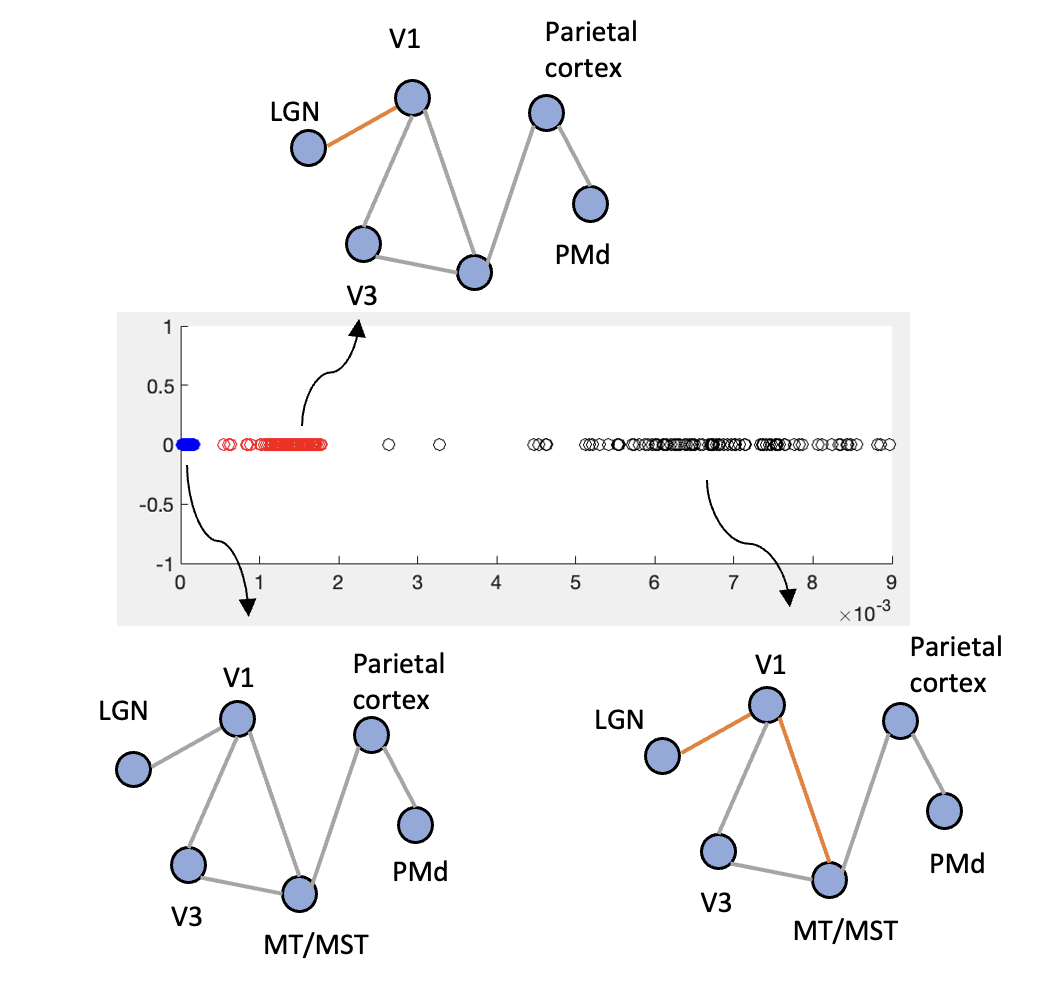}
            \includegraphics[scale=0.28]{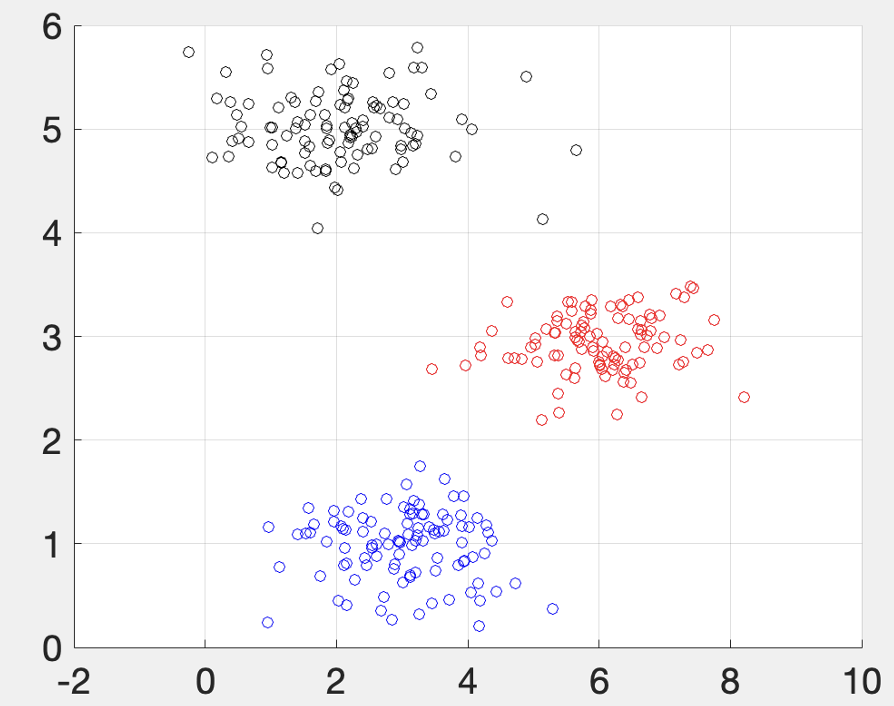}
	\end{center}
	\caption{The figure on the left depicts the output of PMd and some captured important synaptic architecture according to different clusters. The figures on the right illustrate the results obtained by applying our proposed model to segregate the data points into three distinct clusters, represented by the colors black, red and purple, respectively. %The blue dots at the bottom exhibit the output generated by the neural network.
 }
	\label{fig:cluster}
\end{figure}

Next, we evaluate the performance of our proposed model on a simple dataset comprising 300 unlabeled two-dimensional data points for the clustering analysis task, which can be found in \cite{ps3}. The two-dimensional inputs were fed into the SC and LGN modules as constant stimulations one by one, while PMd was treated as the sole output channel. The registered output values at the PMd suggested the data should be grouped into three clusters. It provides a visual correlation between the PMd's output and the underlying synaptic architecture, emphasizing the heterogeneity in connection strengths across different clusters as shown in Fig. \ref{fig:cluster}. It provides a visual correlation between the PMd's output and the underlying synaptic architecture, emphasizing the heterogeneity in connection strengths across different clusters. The results indicate that although the model relies exclusively on local information for weight updates, the learning process can still be effectively realized, demonstrating the system's capability to learn separating hyperplanes by exploring the boundaries of the polytope defined by the synaptic weight constraints. It can also be viewed as analogous to how the human visual system receives information from the environment, encodes it in the primary visual cortex, and classifies different cues into various categories based on the output of neurons in other parts of the visual cortex. Meanwhile, the synaptic weights update gradually according to this information, thereby influencing subsequent outputs. Despite the simplicity of this approach and the lack of explicit supervision, the model exhibits remarkable clustering performance, as evidenced by the separation among black, purple and red dots in Fig. \ref{fig:cluster}.

\begin{comment}
	\begin{figure}[h]
		\begin{center}
			\includegraphics[scale=0.23]{state2.png}
			\includegraphics[scale=0.2]{weight2.png}
		\end{center}
		\caption{(a) Simulation of the macaque monkey network reaches desired states under optimal control solved via Riccati equation. (b) Synaptic weight dynamics evolving over 30 seconds.}
		\label{fig:optimalcontrol2}
	\end{figure}
\end{comment}
%One stem originates from SC, traverse mediodorsal nucleus (MD),frontal eye fields (FEF),prefrontal cortex (PF), inferior temporal cortex (IT) and finally reaching the extremities of posterior division of area IT (TEO). Another stem starts from LGN, traversing through V1, V3, MT/MST, and parietal cortex (comprised of subregions AIP, VIP, LIP, MIP), reaching the extremities of premotor dorsal (PMd)Primary visual areas V1, V2, V4 and pulvinar nucleus (PL) constitute one bud bidirectionally connected to the stem with V1 as the joint. It should be noted that there exist various schemes for partitioning this brain network graph into a generalized cactus representation. Different partitions may represent achieve various tasks.

\section{Conclusions}
In this work, we presented a neuromimetic dynamic network model that captures neural features like synaptic plasticity and Hebbian learning.
%principles with multiple control and outout nodes. Through theoretical proofs and simulations, it exhibits a a good deal of neural features with external inputs. Even though controllability and observability of a complex system are hard to be obtained when the dimension is large, they own generic properties in respect to the determinant of the controllability matrix. We explored the structural controllability and observability regarding the graph structure and came to the conclusion that if the network has a generalized cactus, the system is structural controllability and observability. A macaque monkey vision system whose structure is a generalized cactus can validate our findings. As an avenue for future work, we aim to advance further the systems proposed herein for practical applications. Specifically, we plan to explore how the graph theory, together with the control theory framework developed in this paper, could be leveraged to solve real-world tasks. This will allow assessment of the theoretical results in an applied context and potentially yield enhanced network control and resilience.
While the models we've introduced are fundamentally nonlinear, the analysis has relied on extending the original linear concept of structural controllability to networked control systems with generalized cactus graph topologies.  A stylized model of the macaque visual cortex was proposed and it was shown that even this simple model was capable of learning to classify data.  Future work will be aimed at models in which the network of interoperating brain regions is larger -- with nodes being more functionally diverse.  We also hope to extend the approach to more nuanced kinds of data classification such as object recognition.  The mechanisms that underly learning in our models remain to be more deeply understood.

\begin{comment}
    
\begin{subequations}
\nonumber
	\begin{align}
		\dot x_{i}(t)&=-c_nx_{i}(t)+\sum_{(j,i)\in\mathcal{E}}a_{ij}(t)x_{j}(t)+ b_iu_{i},\\
		a_{ij}(t)&=\left\{\begin{matrix}
			\left[c_a^-a_{ij}(\tau p)+\phi(x_i(\tau p)x_j(\tau p))\right]_{\underline a^-}^{\overline a^-},(j,i)\in\mathcal{E}^-\\
			\left[c_a^+a_{ij}(\tau p)+\phi(x_i(\tau p)x_j(\tau p))\right]_{\underline a^+}^{\overline a^+},
   (j,i)\in\mathcal{E}^+
		\end{matrix}\right. 
	\end{align}
\end{subequations}

\end{comment}
\bibliography{autosam}

\begin{thebibliography}{29}
\expandafter\ifx\csname natexlab\endcsname\relax\def\natexlab#1{#1}\fi
\providecommand{\url}[1]{\texttt{#1}}
\providecommand{\href}[2]{#2}
\providecommand{\path}[1]{#1}
\providecommand{\DOIprefix}{doi:}
\providecommand{\ArXivprefix}{arXiv:}
\providecommand{\URLprefix}{URL: }
\providecommand{\Pubmedprefix}{pmid:}
\providecommand{\doi}[1]{\href{http://dx.doi.org/#1}{\path{#1}}}
\providecommand{\Pubmed}[1]{\href{pmid:#1}{\path{#1}}}
\providecommand{\bibinfo}[2]{#2}
\ifx\xfnm\relax \def\xfnm[#1]{\unskip,\space#1}\fi
%Type = Article
\bibitem[{Agosta et~al.(2013)Agosta, Galantucci, Canu, Cappa, Magnani, Franceschi, Falini, Comi and Filippi}]{agosta2013disruption}
\bibinfo{author}{Agosta, F.}, \bibinfo{author}{Galantucci, S.}, \bibinfo{author}{Canu, E.}, \bibinfo{author}{Cappa, S.F.}, \bibinfo{author}{Magnani, G.}, \bibinfo{author}{Franceschi, M.}, \bibinfo{author}{Falini, A.}, \bibinfo{author}{Comi, G.}, \bibinfo{author}{Filippi, M.}, \bibinfo{year}{2013}.
\newblock \bibinfo{title}{Disruption of structural connectivity along the dorsal and ventral language pathways in patients with nonfluent and semantic variant primary progressive aphasia: a dt mri study and a literature review}.
\newblock \bibinfo{journal}{Brain and language} \bibinfo{volume}{127}, \bibinfo{pages}{157--166}.
%Type = Article
\bibitem[{Arcak(2011)}]{arcak2011diagonal}
\bibinfo{author}{Arcak, M.}, \bibinfo{year}{2011}.
\newblock \bibinfo{title}{Diagonal stability on cactus graphs and application to network stability analysis}.
\newblock \bibinfo{journal}{IEEE Transactions on Automatic Control} \bibinfo{volume}{56}, \bibinfo{pages}{2766--2777}.
%Type = Inproceedings
\bibitem[{Baillieul(2019)}]{baillieul2019perceptual}
\bibinfo{author}{Baillieul, J.}, \bibinfo{year}{2019}.
\newblock \bibinfo{title}{Perceptual control with large feature and actuator networks}, in: \bibinfo{booktitle}{2019 IEEE 58th Conference on Decision and Control (CDC)}, \bibinfo{organization}{IEEE}. pp. \bibinfo{pages}{3819--3826}.
%Type = Article
\bibitem[{Beyeler et~al.(2019)Beyeler, Rounds, Carlson, Dutt and Krichmar}]{beyeler2019neural}
\bibinfo{author}{Beyeler, M.}, \bibinfo{author}{Rounds, E.L.}, \bibinfo{author}{Carlson, K.D.}, \bibinfo{author}{Dutt, N.}, \bibinfo{author}{Krichmar, J.L.}, \bibinfo{year}{2019}.
\newblock \bibinfo{title}{Neural correlates of sparse coding and dimensionality reduction}.
\newblock \bibinfo{journal}{PLoS computational biology} \bibinfo{volume}{15}, \bibinfo{pages}{e1006908}.
%Type = Article
\bibitem[{Burkitt(2006)}]{burkitt2006review}
\bibinfo{author}{Burkitt, A.N.}, \bibinfo{year}{2006}.
\newblock \bibinfo{title}{A review of the integrate-and-fire neuron model: I. homogeneous synaptic input}.
\newblock \bibinfo{journal}{Biological cybernetics} \bibinfo{volume}{95}, \bibinfo{pages}{1--19}.
%Type = Article
\bibitem[{Carvalho et~al.(2017)Carvalho, Pequito, Aguiar, Kar and Johansson}]{carvalho2017composability}
\bibinfo{author}{Carvalho, J.F.}, \bibinfo{author}{Pequito, S.}, \bibinfo{author}{Aguiar, A.P.}, \bibinfo{author}{Kar, S.}, \bibinfo{author}{Johansson, K.H.}, \bibinfo{year}{2017}.
\newblock \bibinfo{title}{Composability and controllability of structural linear time-invariant systems: Distributed verification}.
\newblock \bibinfo{journal}{Automatica} \bibinfo{volume}{78}, \bibinfo{pages}{123--134}.
%Type = Article
\bibitem[{Centorrino et~al.(2024)Centorrino, Bullo and Russo}]{centorrino2024modeling}
\bibinfo{author}{Centorrino, V.}, \bibinfo{author}{Bullo, F.}, \bibinfo{author}{Russo, G.}, \bibinfo{year}{2024}.
\newblock \bibinfo{title}{Modeling and contractivity of neural-synaptic networks with hebbian learning}.
\newblock \bibinfo{journal}{Automatica} \bibinfo{volume}{164}, \bibinfo{pages}{111636}.
%Type = Article
\bibitem[{Citri and Malenka(2008)}]{citri2008synaptic}
\bibinfo{author}{Citri, A.}, \bibinfo{author}{Malenka, R.C.}, \bibinfo{year}{2008}.
\newblock \bibinfo{title}{Synaptic plasticity: multiple forms, functions, and mechanisms}.
\newblock \bibinfo{journal}{Neuropsychopharmacology} \bibinfo{volume}{33}, \bibinfo{pages}{18--41}.
%Type = Article
\bibitem[{Cook et~al.(2019)Cook, Jarrell, Brittin, Wang, Bloniarz, Yakovlev, Nguyen, Tang, Bayer, Duerr et~al.}]{cook2019whole}
\bibinfo{author}{Cook, S.J.}, \bibinfo{author}{Jarrell, T.A.}, \bibinfo{author}{Brittin, C.A.}, \bibinfo{author}{Wang, Y.}, \bibinfo{author}{Bloniarz, A.E.}, \bibinfo{author}{Yakovlev, M.A.}, \bibinfo{author}{Nguyen, K.C.}, \bibinfo{author}{Tang, L.T.H.}, \bibinfo{author}{Bayer, E.A.}, \bibinfo{author}{Duerr, J.S.}, et~al., \bibinfo{year}{2019}.
\newblock \bibinfo{title}{Whole-animal connectomes of both caenorhabditis elegans sexes}.
\newblock \bibinfo{journal}{Nature} \bibinfo{volume}{571}, \bibinfo{pages}{63--71}.
%Type = Article
\bibitem[{Dion et~al.(2003)Dion, Commault and Van~der Woude}]{dion2003generic}
\bibinfo{author}{Dion, J.M.}, \bibinfo{author}{Commault, C.}, \bibinfo{author}{Van~der Woude, J.}, \bibinfo{year}{2003}.
\newblock \bibinfo{title}{Generic properties and control of linear structured systems: a survey}.
\newblock \bibinfo{journal}{Automatica} \bibinfo{volume}{39}, \bibinfo{pages}{1125--1144}.
%Type = Article
\bibitem[{Gal{\'a}n(2008)}]{galan2008network}
\bibinfo{author}{Gal{\'a}n, R.F.}, \bibinfo{year}{2008}.
\newblock \bibinfo{title}{On how network architecture determines the dominant patterns of spontaneous neural activity}.
\newblock \bibinfo{journal}{PloS one} \bibinfo{volume}{3}, \bibinfo{pages}{e2148}.
%Type = Article
\bibitem[{Galtier et~al.(2012)Galtier, Faugeras and Bressloff}]{galtier2012hebbian}
\bibinfo{author}{Galtier, M.N.}, \bibinfo{author}{Faugeras, O.D.}, \bibinfo{author}{Bressloff, P.C.}, \bibinfo{year}{2012}.
\newblock \bibinfo{title}{Hebbian learning of recurrent connections: a geometrical perspective}.
\newblock \bibinfo{journal}{Neural computation} \bibinfo{volume}{24}, \bibinfo{pages}{2346--2383}.
%Type = Book
\bibitem[{Gauthier et~al.(2010)Gauthier, Tarr and Bub}]{gauthier2010perceptual}
\bibinfo{author}{Gauthier, I.}, \bibinfo{author}{Tarr, M.}, \bibinfo{author}{Bub, D.}, \bibinfo{year}{2010}.
\newblock \bibinfo{title}{Perceptual expertise: Bridging brain and behavior}.
\newblock \bibinfo{publisher}{Oxford University Press, New York}.
%Type = Article
\bibitem[{Herculano-Houzel(2009)}]{herculano2009human}
\bibinfo{author}{Herculano-Houzel, S.}, \bibinfo{year}{2009}.
\newblock \bibinfo{title}{The human brain in numbers: a linearly scaled-up primate brain}.
\newblock \bibinfo{journal}{Frontiers in human neuroscience} \bibinfo{volume}{3}, \bibinfo{pages}{857}.
%Type = Book
\bibitem[{Kandel et~al.(2000)Kandel, Schwartz, Jessell, Siegelbaum, Hudspeth, Mack et~al.}]{kandel2000principles}
\bibinfo{author}{Kandel, E.R.}, \bibinfo{author}{Schwartz, J.H.}, \bibinfo{author}{Jessell, T.M.}, \bibinfo{author}{Siegelbaum, S.}, \bibinfo{author}{Hudspeth, A.J.}, \bibinfo{author}{Mack, S.}, et~al., \bibinfo{year}{2000}.
\newblock \bibinfo{title}{Principles of neural science}. volume~\bibinfo{volume}{4}.
\newblock \bibinfo{publisher}{McGraw-hill New York}.
%Type = Article
\bibitem[{Kim and Bassett(2020)}]{kim2020linear}
\bibinfo{author}{Kim, J.Z.}, \bibinfo{author}{Bassett, D.S.}, \bibinfo{year}{2020}.
\newblock \bibinfo{title}{Linear dynamics and control of brain networks}.
\newblock \bibinfo{journal}{Neural Engineering} , \bibinfo{pages}{497--518}.
%Type = Article
\bibitem[{Krotov and Hopfield(2019)}]{krotov2019unsupervised}
\bibinfo{author}{Krotov, D.}, \bibinfo{author}{Hopfield, J.J.}, \bibinfo{year}{2019}.
\newblock \bibinfo{title}{Unsupervised learning by competing hidden units}.
\newblock \bibinfo{journal}{Proceedings of the National Academy of Sciences} \bibinfo{volume}{116}, \bibinfo{pages}{7723--7731}.
%Type = Article
\bibitem[{Li et~al.(1996)Li, Xi and Zhang}]{li1996cactus}
\bibinfo{author}{Li, K.}, \bibinfo{author}{Xi, Y.}, \bibinfo{author}{Zhang, Z.}, \bibinfo{year}{1996}.
\newblock \bibinfo{title}{Cactus-like and structural controllability of interconnected dynamical system}.
\newblock \bibinfo{journal}{IFAC Proceedings Volumes} \bibinfo{volume}{29}, \bibinfo{pages}{4452--4457}.
%Type = Article
\bibitem[{Lillicrap et~al.(2020)Lillicrap, Santoro, Marris, Akerman and Hinton}]{lillicrap2020backpropagation}
\bibinfo{author}{Lillicrap, T.P.}, \bibinfo{author}{Santoro, A.}, \bibinfo{author}{Marris, L.}, \bibinfo{author}{Akerman, C.J.}, \bibinfo{author}{Hinton, G.}, \bibinfo{year}{2020}.
\newblock \bibinfo{title}{Backpropagation and the brain}.
\newblock \bibinfo{journal}{Nature Reviews Neuroscience} \bibinfo{volume}{21}, \bibinfo{pages}{335--346}.
%Type = Article
\bibitem[{Lin(1974)}]{lin1974structural}
\bibinfo{author}{Lin, C.T.}, \bibinfo{year}{1974}.
\newblock \bibinfo{title}{Structural controllability}.
\newblock \bibinfo{journal}{IEEE Transactions on Automatic Control} \bibinfo{volume}{19}, \bibinfo{pages}{201--208}.
%Type = Article
\bibitem[{Mayeda(1981)}]{mayeda1981structural}
\bibinfo{author}{Mayeda, H.}, \bibinfo{year}{1981}.
\newblock \bibinfo{title}{On structural controllability theorem}.
\newblock \bibinfo{journal}{IEEE Transactions on Automatic Control} \bibinfo{volume}{26}, \bibinfo{pages}{795--798}.
%Type = Article
\bibitem[{Menara et~al.(2018)Menara, Bassett and Pasqualetti}]{menara2018structural}
\bibinfo{author}{Menara, T.}, \bibinfo{author}{Bassett, D.S.}, \bibinfo{author}{Pasqualetti, F.}, \bibinfo{year}{2018}.
\newblock \bibinfo{title}{Structural controllability of symmetric networks}.
\newblock \bibinfo{journal}{IEEE Transactions on Automatic Control} \bibinfo{volume}{64}, \bibinfo{pages}{3740--3747}.
%Type = Article
\bibitem[{Michel et~al.(2022)Michel, Jha and Ewetz}]{michel2022survey}
\bibinfo{author}{Michel, A.}, \bibinfo{author}{Jha, S.K.}, \bibinfo{author}{Ewetz, R.}, \bibinfo{year}{2022}.
\newblock \bibinfo{title}{A survey on the vulnerability of deep neural networks against adversarial attacks}.
\newblock \bibinfo{journal}{Progress in Artificial Intelligence} , \bibinfo{pages}{1--11}.
%Type = Article
\bibitem[{Muldoon et~al.(2016)Muldoon, Pasqualetti, Gu, Cieslak, Grafton, Vettel and Bassett}]{muldoon2016stimulation}
\bibinfo{author}{Muldoon, S.F.}, \bibinfo{author}{Pasqualetti, F.}, \bibinfo{author}{Gu, S.}, \bibinfo{author}{Cieslak, M.}, \bibinfo{author}{Grafton, S.T.}, \bibinfo{author}{Vettel, J.M.}, \bibinfo{author}{Bassett, D.S.}, \bibinfo{year}{2016}.
\newblock \bibinfo{title}{Stimulation-based control of dynamic brain networks}.
\newblock \bibinfo{journal}{PLoS computational biology} \bibinfo{volume}{12}, \bibinfo{pages}{e1005076}.
%Type = Article
\bibitem[{Pehlevan et~al.(2015)Pehlevan, Hu and Chklovskii}]{pehlevan2015hebbian}
\bibinfo{author}{Pehlevan, C.}, \bibinfo{author}{Hu, T.}, \bibinfo{author}{Chklovskii, D.B.}, \bibinfo{year}{2015}.
\newblock \bibinfo{title}{A hebbian/anti-hebbian neural network for linear subspace learning: A derivation from multidimensional scaling of streaming data}.
\newblock \bibinfo{journal}{Neural computation} \bibinfo{volume}{27}, \bibinfo{pages}{1461--1495}.
%Type = Misc
\bibitem[{Plummer and Saenko(2022)}]{ps3}
\bibinfo{author}{Plummer, B.}, \bibinfo{author}{Saenko, K.}, \bibinfo{year}{2022}.
\newblock \bibinfo{title}{{BU CS542 ps3}}.
\newblock \bibinfo{howpublished}{\url{https://github.com/zexins/BU-CS542-ps3/blob/main/data2.mat}}.
\newblock \bibinfo{note}{[Online; accessed 17-June-2024]}.
%Type = Article
\bibitem[{Rech and Perret(1990)}]{rech1990structural}
\bibinfo{author}{Rech, C.}, \bibinfo{author}{Perret, R.}, \bibinfo{year}{1990}.
\newblock \bibinfo{title}{Structural observability of interconnected systems}.
\newblock \bibinfo{journal}{International journal of systems science} \bibinfo{volume}{21}, \bibinfo{pages}{1881--1888}.
%Type = Inproceedings
\bibitem[{Sun and Baillieul(2022)}]{sun2022neuromimetic}
\bibinfo{author}{Sun, Z.}, \bibinfo{author}{Baillieul, J.}, \bibinfo{year}{2022}.
\newblock \bibinfo{title}{Neuromimetic linear systems -- resilience and learning}, in: \bibinfo{booktitle}{2022 IEEE 61st Conference on Decision and Control (CDC)}, \bibinfo{organization}{IEEE}. pp. \bibinfo{pages}{7388--7394}.
%Type = Inproceedings
\bibitem[{Sun and Baillieul(2024)}]{sun2024neuromimetic}
\bibinfo{author}{Sun, Z.}, \bibinfo{author}{Baillieul, J.}, \bibinfo{year}{2024}.
\newblock \bibinfo{title}{Neuromimetic dynamic networks with hebbian learning}, in: \bibinfo{booktitle}{2024 American Control Conference (ACC)}, \bibinfo{organization}{IEEE}. pp. \bibinfo{pages}{5264--5269}.

\end{thebibliography}

\end{document}